\documentclass[manuscript,screen,acmsmall]{acmart}

\DeclareRobustCommand\EEOClongname{}
\DeclareRobustCommand\UNDPlongname{}

\usepackage{bbm}
\usepackage{array}
\usepackage[normalem]{ulem}
\useunder{\uline}{\ul}{}
\usepackage{longtable}
\usepackage{threeparttablex}
\usepackage{enumitem}
\usepackage[framemethod=tikz]{mdframed}

\definecolor[named]{teal}{cmyk}{0.80, 0.20, 0.40, 0.20}
\definecolor[named]{revcolor}{cmyk}{1.00, 0.10, 0.0, 0.10}
\definecolor[named]{violetred}{cmyk}{0, 0.75, 0.25, 0.20}
\definecolor[named]{darkgreen}{cmyk}{1, 0, 1, 0.50}

\newcolumntype{x}[1]{>{\centering\arraybackslash\hspace{0pt}}p{#1}}

\newif\ifdraft
\draftfalse
\newcommand{\todo}[1]{\ifdraft{\textcolor{ACMDarkBlue}{[TODO]: {#1}} }\else{\vspace{0ex}}\fi}

\newif\ifrevision
\revisionfalse
\newcommand{\wasnew}[1]{{#1}}
\newcommand{\wasremove}[1]{{\vspace{0ex}}}
\newcommand{\wasreplace}[2]{{#2}}

\newcommand{\new}[1]{\ifrevision{\textcolor{revcolor}{{#1}} }\else{{#1}}\fi}

\newcommand{\replace}[2]{\ifrevision{\textcolor{revcolor}{\sout{#1}}\textcolor{revcolor}{{#2}}}\else{{#2}}\fi}

\newcommand{\stcomp}[1]{{#1}^{\mathsf{c}}}
\newcommand{\tpr}[0]{\text{TPR}}
\newcommand{\tnr}[0]{\text{TNR}}

\newcommand{\fnr}[0]{\text{FNR}}
\newcommand{\maxg}[0]{\max_{g \in \mathcal{S}}}
\newcommand{\ming}[0]{\min_{g \in \mathcal{S}}}
\newmdenv[innerlinewidth=0.5pt, roundcorner=4pt,linecolor=red,innerleftmargin=6pt,
innerrightmargin=6pt,innertopmargin=6pt,innerbottommargin=6pt]{mybox}

\DeclareMathOperator{\di}{DI}
\DeclareMathOperator{\lrr}{LRR}
\DeclareMathOperator{\ndkl}{NDKL}
\DeclareMathOperator{\skewk}{\text{skew@}k}
\DeclareMathOperator{\rpp}{RPP}
\DeclareMathOperator{\dd}{DD}
\DeclareMathOperator{\tprd}{TPRD}
\DeclareMathOperator{\rms}{RMS}
\DeclareMathOperator{\xeo}{xEO}
\DeclareMathOperator{\fnrr}{FNRR}
\DeclareMathOperator{\bcrd}{BCRD}
\DeclareMathOperator{\midif}{MID}
\DeclareMathOperator{\sd}{SD}
\DeclareMathOperator{\sauc}{sAUC}
\DeclareMathOperator{\gbs}{GBS}
\DeclareMathOperator{\mia}{MIA}
\DeclareMathOperator{\gtr}{GTR}
\DeclareMathOperator{\drd}{DRD}
\DeclareMathOperator{\med}{MED}
\DeclareMathOperator{\mae}{MAE}
\DeclareMathOperator{\skl}{SKL}
\newcommand{\no}[0]{No}

\newcommand{\namedpar}[1]{\vspace{0.1 cm} \noindent \textbf{#1}}

\AtBeginDocument{%
  \providecommand\BibTeX{{%
    \normalfont B\kern-0.5em{\scshape i\kern-0.25em b}\kern-0.8em\TeX}}}

\setcopyright{rightsretained}
\acmJournal{TIST}
\acmYear{2025} \acmVolume{1} \acmNumber{1} \acmArticle{1} \acmMonth{1}\acmDOI{10.1145/3696457}

\acmConference[ACM]{Make sure to enter the correct
  conference title from your rights confirmation emai}{January 2025}{}
\acmPrice{15.00}
\acmISBN{978-1-4503-XXXX-X/18/06}

\begin{document}

\title{Fairness and Bias in Algorithmic Hiring: \\ A Multidisciplinary Survey}


\author{Alessandro Fabris}
\affiliation{%
  \institution{Max Planck Institute for Security and Privacy}
  \city{Bochum}
  \country{Germany}}
\email{alessandro.fabris@mpi-sp.org}

\author{Nina Baranowska}
\authornote{Authors listed in alphabetical order.}
\affiliation{%
  \institution{Radboud University}
  \city{Nijmegen}
  \country{The Netherlands}}
\email{nina.baranowska@ru.nl}

\author{Matthew J. Dennis}
\authornotemark[1]
\affiliation{%
  \institution{Eindhoven University of Technology}
  \city{Eindhoven}
  \country{The Netherlands}}
\email{m.j.dennis@tue.nl}

\author{David Graus}
\authornotemark[1]
\affiliation{%
  \institution{Randstad}
  \city{Diemen}
  \country{The Netherlands}}
\email{david.graus@randstad.com}

\author{Philipp Hacker}
\authornotemark[1]
\affiliation{%
  \institution{European University Viadrina}
  \city{Frankfurt (Oder)}
  \country{Germany}}
\email{hacker@europa-uni.de}

\author{Jorge Saldivar}
\authornotemark[1]
\affiliation{%
  \institution{Universitat Pompeu Fabra}
  \city{Barcelona}
  \country{Spain}}
\email{jorge.saldivar@upf.edu}

\author{Frederik Zuiderveen Borgesius}
\authornotemark[1]
\affiliation{%
  \institution{Radboud University}
  \city{Nijmegen}
  \country{The Netherlands}}
\email{frederikzb@cs.ru.nl}

\author{Asia J. Biega}
\affiliation{%
  \institution{Max Planck Institute for Security and Privacy}
  \city{Bochum}
  \country{Germany}}
\email{asia.biega@mpi-sp.org}

\renewcommand{\shortauthors}{Fabris et al.}

\begin{abstract}
  Employers are adopting algorithmic hiring technology throughout the recruitment pipeline. Algorithmic fairness is especially applicable in this domain due to its high stakes and structural inequalities. Unfortunately, most work in this space provides partial treatment, often constrained by two competing narratives, optimistically focused on replacing biased recruiter decisions or pessimistically pointing to the automation of discrimination. Whether, and more importantly \emph{what types of}, algorithmic hiring can be less biased and more beneficial to society than low-tech alternatives currently remains unanswered, to the detriment of trustworthiness. This multidisciplinary survey caters to practitioners and researchers with a balanced and integrated coverage of systems, biases, measures, mitigation strategies, datasets, and legal aspects of algorithmic hiring and fairness. Our work supports a contextualized understanding and governance of this technology by highlighting current opportunities and limitations, providing recommendations for future work to ensure shared benefits for all stakeholders.
\end{abstract}

\ccsdesc[500]{General and reference~Surveys and overviews}
\ccsdesc[300]{Information systems~Data management systems}
\ccsdesc[500]{Social and professional topics~Computing / technology policy}

\keywords{Algorithmic hiring, Online recruitment, Algorithmic fairness, Bias, Anti-discrimination}

\received{07 September 2023}


\maketitle

\begin{mybox}
\textbf{Correct reference for this work}:

\vspace{0.2cm}
\begin{small}
\noindent Alessandro Fabris, Nina Baranowska, Matthew J. Dennis, David Graus, Philipp Hacker, Jorge Saldivar, Frederik Zuiderveen Borgesius, and Asia J. Biega. Fairness and Bias in Algorithmic Hiring: a Multidisciplinary Survey. ACM Transactions on Intelligent Systems and Technology. 2025.

\noindent \url{https://doi.org/10.1145/3696457} 
\end{small}
\end{mybox}


\section{Introduction}
\label{sec:intro}

\todo{If we want to update, a few recent articles are \citet{sogancioglu2023using} and the ones from HR recsys.}

New algorithms for Human Resources (HR) are developed and deployed every year. By one count, there are over 250 Artificial Intelligence (AI) tools for HR on the market \citep{wef2021human}, with entire manuals for HR professionals available on the topic \citep{eubanks2022artificial}. The average job posting yields more than 100 candidates \citep{pwc2017artificial,fuller2021hidden}. These mutually reinforcing factors were accelerated by the COVID\nobreakdash-19 pandemic and recent advances in AI \citep{malik2022may,qin2023towards}. As a result, prospective job applicants and their chances of success are increasingly influenced by algorithmic hiring technology, including automated job descriptions, resume parsers, and video interviews.

Workplaces and labor markets are fraught with biases, imbalances, and patterns of discrimination against vulnerable groups, including women, ethnic minorities, and people with disabilities \citep{azmat2014gender,bertrand2004emily,riach2002field}. While algorithmic hiring represents an opportunity to mitigate these biases, it also runs the risk of reinforcing and amplifying them, causing harm and hindering trustworthiness \citep{liu2023trustworthy}. The debate on this topic is often polarized between techno-enthusiasm \citep{miller2018want,kappen2021objective} and
pessimism \citep{imana2021auditing,andrews2022automating} due to partial perspectives on a field that is large and complex. In this paper, we offer a multidisciplinary survey of fairness and bias in algorithmic hiring centered on computer science (focused on systems, algorithms, metrics, and datasets) and informed by related disciplines. We critically analyze available resources and methods, highlight common challenges, and identify opportunities to advance the field.


\namedpar{Related work}. 
\citet{bogen2018help} presented a technical report that described algorithmic hiring tools available in 2018, together with selected \wasreplace{bias conducive factors (BCFs)}{sources of bias}, and provided a US-centric review of relevant laws and policies. 
\citet{kochling2020discriminated} conducted a \wasreplace{qualitative review}{review of algorithms in HR from a business research perspective,} focused on non-empirical articles. \wasnew{Their qualitative discussion raises awareness on discrimination in HR algorithms without delving into measures and mitigation strategies.}
\wasreplace{Rieskamp et al. summarized nine fairness-enhancing methods for algorithmic hiring. Table [?] offers a comparison of existing overviews, underscoring the opportunity for an integrated survey.}{The prior work most closely aligned to ours is \citet{rieskamp2023approaches}, surveying nine articles on bias mitigation for algorithmic hiring. Our survey expands on this work by presenting more bias mitigation techniques, covering fairness measures, describing available datasets, and by situating them in the broader social and legal context characterizing algorithmic hiring. In concurrent work, \citet{kumar2023fairness}, survey the literature on fair recommender systems in the recruitment domain. While similar in spirit, their work focuses on ranking; our work presents a broader view of algorithmic hiring across different tasks and hiring stages, covering many technologies and bias conducive factors throughout the algorithmic hiring pipeline.}


\namedpar{Contributions and audience}. This work provides a contextualized treatment of fairness and bias in algorithmic hiring. It was carried out by a team with mixed backgrounds in computer science, law, and philosophy, and with input from practitioners developing algorithmic hiring products. Our contributions, catering \wasreplace{to both}{especially to} practitioners and researchers, can roughly be divided as follows. Practitioners \wasnew{such as data scientists, engineers, and product managers,} will find (1) a detailed list and description of domain-specific factors that lead to biases in their systems \wasnew{(Sec. \ref{sec:bcf}),} (2) methods to mitigate these biases \wasnew{(Sec. \ref{sec:miti}), (3) guidance on their applicability in practice (Sec. \ref{sec:practice}),} and \wasreplace{(3)}{(4)} pointers to key legal references in the EU and the US \wasnew{(Sec. \ref{sec:legal}).} Researchers will benefit from (4) an up-to-date description of hiring technology \wasnew{(Sec. \ref{sec:pipeline}),} (5) a unified treatment of fair hiring measures \wasnew{(Sec. \ref{sec:measures}),} and (6) a collection of datasets in this space \wasnew{(Sec. \ref{sec:data}).} Overall, an integrated coverage of these topics provides (7) a gentle primer for readers who are not experts in the field and (8) highlights important gaps and promising directions for future work in computer science at the intersection with law and policy \wasnew{(Sec. \ref{sec:pros_cons}).} 
\wasnew{Considering a broader audience, Sections~\ref{sec:pipeline}, \ref{sec:bcf}, \ref{sec:legal}, \ref{sec:pros_cons}, and \ref{sec:concl} cover a shared background which should be relevant for all readers, including HR professionals and legal scholars.} 

\namedpar{Structure}. The remainder of this work is organized as follows. Section \ref{sec:pipeline} introduces the main stages and systems in the algorithmic hiring pipeline. Section \ref{sec:bcf} focuses on bias, summarizing the most important factors in the labor market, the recruitment sector, and the tech industry that can lead to unfair \wasnew{recruitment} systems.  Sections \ref{sec:measures}--\ref{sec:data} result from a systematic review of the literature with methods described in Appendix \ref{sec:lit_rev}. Sections \ref{sec:measures} and \ref{sec:miti} describe the main fairness measures and mitigation approaches, while Section \ref{sec:data} presents the datasets used in the algorithmic hiring literature. \wasnew{Section \ref{sec:practice} discusses practical aspects of anti-discrimination in algorithmic hiring guiding the choice of fairness measures and mitigation strategies.} Sections \ref{sec:legal} and \ref{sec:messy} widen the perspective beyond computer science by outlining the main legal frameworks and situating algorithmic hiring in its broader socio-technical context. Section \ref{sec:pros_cons} summarizes opportunities, limitations, and recommendations for future work; Section \ref{sec:concl} provides concluding remarks.

\section{The Algorithmic Hiring Pipeline}
\label{sec:pipeline}

\begin{figure}[tb]
  \centering
  \includegraphics[width=1\textwidth]{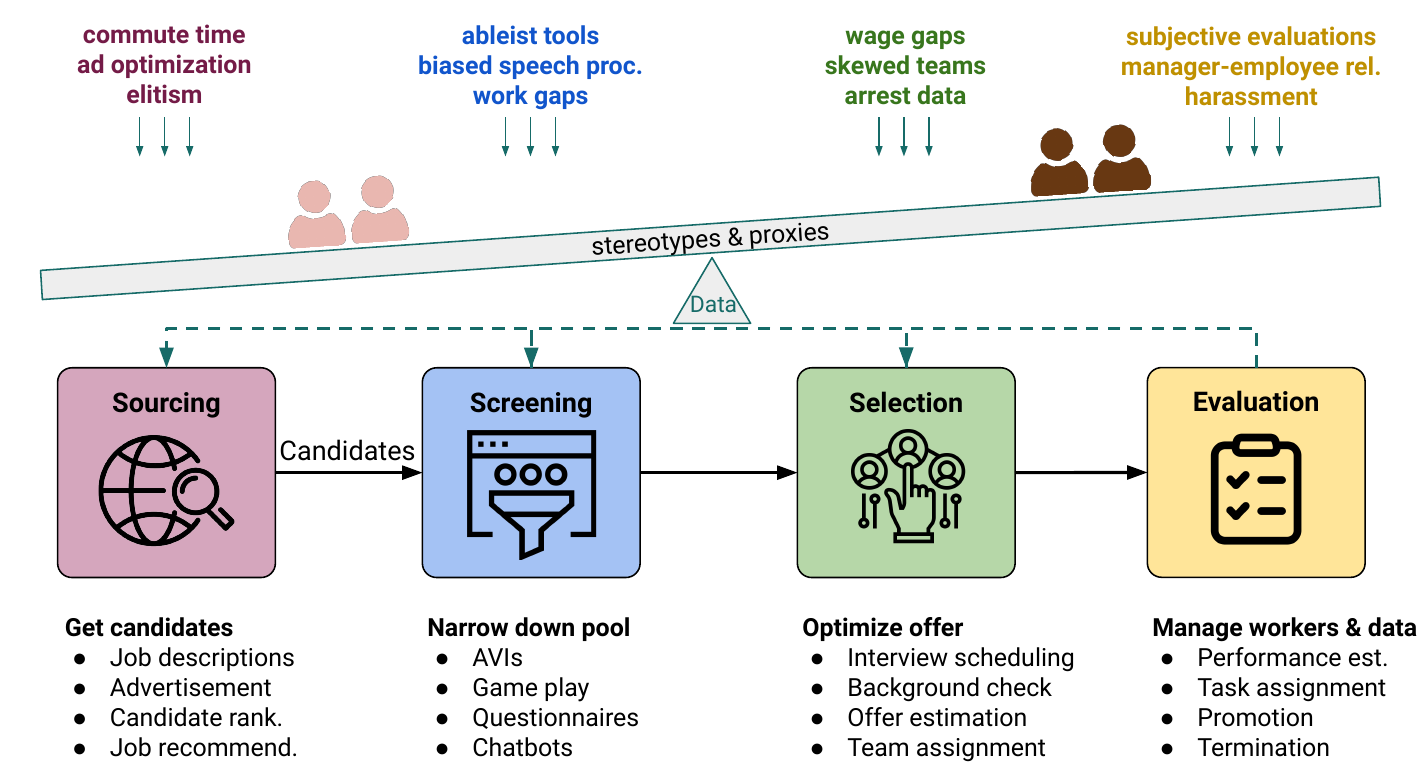}
  \caption{Stages of algorithmic hiring with the main tools (below) and bias conducive factors (above) tilting data against vulnerable groups, jointly with stereotypes and sensitive attribute proxies.}
  \label{fig:summary}
\end{figure}

Algorithmic hiring comprises algorithms, tools, and systems to automate or assist HR decisions on candidate recruitment and evaluation. Elaborating on previous work from civil society \citep{bogen2018help}, regulatory bodies \citep{ep2021proposal}, and academia \citep{raghavan2020mitigating}, we distinguish four stages in the algorithmic hiring pipeline, reported in Figure \ref{fig:summary}.

\subsection*{Sourcing}

The first stage of the hiring pipeline provides employers with a large pool of candidates for a given position. Historically, this is the hiring stage where algorithms are most prominent and well-researched. The main technological solutions and systems are summarized below.

\begin{itemize}
        \item \textbf{Job advertisement} consists of tools to describe a vacancy and make it visible. Job descriptions \citep{hu2022balancing}, possibly written by \wasreplace{neural networks}{language models} \citep{qin2023towards}, are optimized and shared through suitable channels, including employer websites \citep{cerioli2020what}, dedicated platforms \citep{pejicbach2020text}, and ad delivery services \citep{lambrecht2019algorithmic,ali2019discrimination}.
        \item \textbf{Search, ranking, and recommendation} algorithms favor \wasreplace{employee-employer matches}{matches between jobs and job seekers} by ranking candidates for openings \citep{geyik2019fairness} \wasnew{and vice versa:} recommending openings to job seekers \citep{nandy2022achieving}, based on a combination of machine learning, boolean matching, and sorting (e.g. skills, location, industry), supported by information extraction systems \citep{chen2018two}. \wasnew{Fairness is a critical and well-studied property for these systems \citep{wang2023survey,kumar2023fairness}.}
        \item \textbf{Social networks}, especially those embedded in job platforms, play an important role in the visibility of candidates \citep{stoica2018algorithmic} and their awareness of opportunities \citep{fish2019gaps}, \wasnew{favoring professional connections \citep{candela2023disentangling},} contributing to the dissemination of information, and facilitating referral \citep{zhang2021unequal}.
    \end{itemize}

It is worth noting that stage categorization is elastic and usage-dependent. For example, algorithms that extract information from CVs to score employees against job descriptions are central to the sourcing stage, but may also be used for screening \citep{cowgill2018bias}.

\subsection*{Screening}

After sourcing, employers must narrow down large pools of candidates into manageable subsets that can be more thoroughly evaluated by HR specialists and other employees. 

\begin{itemize}
    \item \textbf{Gameplay} is used for psychometric assessment and soft skills measurement. Companies develop proprietary suites of games and make them available to candidates, e.g. through a mobile app. Models analyze their gameplay to explicitly estimate competency scores \citep{orcaa2020description} or implicitly look for similarities with desirable candidates such as current employees \citep{wilson2021building}.
    \item \textbf{Asynchronous Video Interviews (AVI)} are recordings of candidates' answers to a specific set of questions in front of a camera. AVI models rely on different data modalities, including visual (e.g. facial expressions), verbal (e.g. length of sentences) and paraverbal features (e.g. tone) to infer a variety of traits related to personality and hireability \citep{hemamou2021judge,booth2021bias}.
    \item \textbf{Questionnaires} in hiring cover a wide range of purposes, such as personality assessment \citep{hughes2017using}, job performance inference \citep{fernandez2019assessing,burke2021fair} or explicitly asking information about job requirements to filter candidates (e.g. Driver's license ownership) or prioritize them (e.g. years of experience).
    \item \textbf{\wasreplace{Chat bots}{Chatbots}} can mediate interactions between employers and employees by asking basic screening questions for job seekers and scheduling interviews \citep{chandler2018ai,bogen2018help}. Advancements in large language models \citep{ouyang2022training} are likely to increase the influence of this technology on hiring processes, broadening its reach to other stages of recruitment, including sourcing and evaluation \citep{malik2022may,salinas2023unequal}. 
\end{itemize}

\subsection*{Selection}

After screening, candidates are interviewed by HR specialists or other employees. These interviews can include technical and cultural questions influenced by previous stages, e.g. by questionnaire answers. Several tools are available at this stage to help employers select the most desirable job seekers and extend them a suitable offer.

\begin{itemize}
    \item \textbf{Compensation and benefit optimization} allows employers to target their offers to each candidate, thanks to tools that estimate the likelihood of acceptance based on salary, bonus, stock options, and other benefits \citep{bogen2018help,meng2022fine}.
    \item \textbf{Background checks}, primarily focused on criminal records \citep{denver2020criminal,craigie2020ban,lageson2021digitizing}, are used by employers to obtain additional information on candidates before hiring them. Background checks can also target social media \citep{hosain2021social}.
    \item \textbf{Team placement} concerns the assignment of a selected candidate to a specific role within a team or project \citep{pessach2020employees,gomezzara2020taxonomy}.
\end{itemize}

\subsection*{Evaluation}

The hired employees are managed and evaluated on their career progression. While, strictly speaking, these are post-hiring processes, it is fundamental to consider them due to their feedback loops on algorithmic recruitment.  

\begin{itemize}
    \item \textbf{Performance and career management} tools facilitate employee development and monitoring. Technology in this area supports training assignment \citep{budhwar2022artificial}, work allocation \citep{vallance2023tuc}, career development \citep{oracle2023welcome}, along with monitoring and estimation through increasingly available tracking technology \citep{kaur2017game,sonderling2022promise,mulligan2023hearing,skopeliti2023feel}.
    \item \textbf{Engagement analytics} measures employee satisfaction, commitment, and retention probability, often leveraging dedicated questionnaires and surveys \citep{kochling2020discriminated,yaneva2018employee}.  
    \item \textbf{Career progression} deals with promotion and dismissal of employees. Succession plans and promotions can be partially automated \citep{oracle2023welcome,long2018prediction}. Turnover prediction is an active area of research, typically presented as part of analytics to favor retention \citep{punnoose2016prediction}. It is worth noting that turnover data, frequently used to improve hiring systems \citep{burke2021fair}, can also be repurposed towards employee termination. Indeed 98\% of 300 HR leaders from US companies interviewed in a 2023 survey, report that software and algorithms will assist them with layoff decisions throughout the year \citep{verma2023ai}.
\end{itemize}

The progression of candidates through the stages of the hiring pipeline is accompanied by a feedback loop in which the decisions at each stage generate data that influence the remaining stages in subsequent interactions. Evaluation, for example, can lead to job termination impacting tenure, which, in turn, represents a key desideratum and prediction target for sourcing and screening algorithms. These data are influenced by a diverse and complex set of factors, many of which can lead to undesirable discrimination, as described in the next section. 

\section{Bias Conducive Factors}
\label{sec:bcf}

Hiring data and algorithms can display undesirable groupwise patterns caused by Bias Conducive Factors (BCFs) affecting the employment domain. These patterns put some job seekers at a systematic disadvantage based on sensitive attributes such as age, disability, gender, religion or belief, racial or ethnic origin, or sexual orientation. Overall, disparities may exist  in data sample composition and feature values across sensitive groups that reflect and amplify problematic structural differences in society. First, hiring datasets display \wasreplace{differential measurement errors}{measurement errors whose severity varies with protected groups}, most critically in target variables related to employability, reflecting current biases in human ratings. Moreover, biased decisions at early stages result in downstream samples misrepresenting certain groups. Finally, there are some higher-order effects caused by technological blindspots and biases in external tools integrated into the hiring pipeline. Some of these biases are reported in Figure \ref{fig:summary}; at their root there are two overarching BCFs that are worth highlighting from the outset. 

\namedpar{Stereotypes} are widely held beliefs about groups of individuals with a common trait, including their propensity and ability to perform a given job \citep{heilman2012gender}. Stereotypes are sustained by culture, socialization, and experience \citep{cheryan2015cultural,cheryan2015cultural}. They are often acquired at an early age \citep{jaxon2019acquisition} and can be activated unconsciously \citep{moskowitz2012implicit}. Even when outspokenly rejected, stereotypes affect the lives of individuals both descriptively and prescriptively based on coarse categories \citep{ellemers2018:gs}. In turn, this ends up shaping expectations about the qualities, priorities, and needs that people have about themselves and others, including, and perhaps especially, about work \citep{bobbitt2011:gd,cheryan2015cultural,heilman2012gender,undp2023breaking}. For example, \emph{agency} (orientation towards leadership and goal attainment) is stereotypically associated with men, while communion (warmth and propensity for care) is frequently associated with women \citep{eagly2019:gs,heilman2012gender}, with far-reaching effects of gender roles and expectations in employment \citep{heilman2012gender,riach2002field}.

\namedpar{Sensitive attribute proxies.} 
Stereotype activation and entrenchment is not always direct; stereotypes about a group can be activated by proxies \citep{reyna2009blame}. This is especially true for algorithms trained to make inferences from inputs that are strongly correlated with sensitive attributes. For instance, video interviews and resumes contain a wealth of information on gender, race, and other sensitive attributes  \citep{deshpande2020mitigating,dearteaga2019bias,rus2022closing,rivera2011ivies,orcaa2020description}, which can lead algorithms to learn stereotypical associations between sensitive and target variables encoded in the data. More specifically, voice timbre and physical appearance in videos may be used as proxies for gender, while names and spoken languages in CVs may correlate heavily with certain migration backgrounds. Sensitive attribute proxies allow models to learn and reflect the diverse BCFs described in the following section.

\subsection{Institutional biases}
\label{sec:inst_bias}

The first family of BCFs we introduce are institutional biases. These are practices, habits, and norms shared at institutions, such as companies and societies, which reflect negatively on the probability of positive algorithmic hiring outcomes for disadvantaged groups.

\namedpar{Direct discrimination} takes place when disparate outcomes are explicitly caused by sensitive attributes. This type of bias can be difficult to prove, as it requires careful control over non-sensitive attributes to keep them as constant as possible, while only varying sensitive attributes. Famous correspondence studies have shown that black applicants are less likely to be contacted after applying for a job compared to otherwise identical white candidates \citep{kline2022systemic,bertrand2004emily}. Similar results have been found in interview evaluations provided by US-based judges, showing a preference for standard American accents over international ones \citep{leong2019are,purkiss2006implicit,deprez2010accents}. Numerous field experiments using matched pairs of applicants have highlighted discrimination against women and non-whites; experiments also point to a risk of direct discrimination against disabled and older applicants \citep{riach2002field,quillian2017meta}. This is the most obvious BCF; a variety of more subtle ones are listed below.

\namedpar{Horizontal segregation}. Job segregation (i.e. the world today) plays a fundamental role in hiring decisions (i.e. the world tomorrow). Prior experience is considered a fundamental predictor of suitability for a given position \citep{fuller2021hidden}. Horizontal segregation concerns differences in employment rate across industry sectors associated with sensitive attributes, such as gender and race \citep{bloksgaard2011masculinities,tesfai2020dimensions}. Strong gender patterns in diverse regions of the world are linked to persistent gender stereotypes about agency and communion that shape our perception and expectations \citep{bls2023,eige2022gender,gradin2021occupational}. Most field experiments, for example, demonstrate discrimination against women in men-dominated jobs and vice versa \citep{riach2002field,rich2014field}. 

\namedpar{Vertical segregation} summarizes differences in career progression to leadership positions. Predominantly analyzed with respect to binary gender \citep{cotter2001glass,eige2020gender}, recent studies have also highlighted \emph{glass ceiling effects} for non-binary workers \citep{davidson2016gender}, racial minorities \citep{hegde2022race}, and intersectional identities \citep{woodhams2015presence}. When translated into data and dignified with a \emph{ground truth} status, vertical segregation leads to models that reinforce wage gaps \citep{rus2022closing} and lack of diversity in high-status positions.

\namedpar{Cultural fit} is often considered predictive of a candidate's ability to conform and
adapt to the core values and collective behaviors of an organization. Evaluations of cultural fit by recruiters, subjective or objective, can contribute to maintaining a uniform workforce in the company, especially in more senior positions, reinforcing horizontal and vertical segregation \citep{rivera2012hiring,ajunwa2019paradox}.

\namedpar{Elitism} in recruiter evaluations favors candidates educated at prestigious institutions who can also list specific extracurricular accomplishments correlated with socioeconomic status \citep{rivera2011ivies}, further entrenching family status and social class \citep{mullen2009elite,chetty2023diversifying}.

\namedpar{Biased employee evaluation} has different forms and derives from multiple causes. Two key drivers are stereotypes and employee-manager relationships. Gender and race stereotypes about competence drive people's perception on the workplace and potentially bias supervisor evaluations against female and black employees \citep{smith2019power,he2019stereotypes,terrell2017gender,sidanius1989job}.  Furthermore, a combination of vertical job segregation and homophily entails that minority employees who are not well represented in management positions may receive lower ratings when manager-employee personal relationships have a positive influence on evaluations and promotions  \citep{kraiger1985meta,bol2011determinants,biggerstaff2023hitting}

\namedpar{Stereotype violation} has a disparate effect on the ``transgressor'', depending on their gender.  Women are frequently penalized for gender norm violations that make them appear more agentic (e.g. competent and self-confident) and less communal (e.g. kind and warm) than stereotypically expected \citep{parasurama2022gendered,he2021covering,tyler2009violating}. This is especially problematic in the hiring domain, where agency is deemed important for leadership positions and it is harder to demonstrate for women without giving an impression of low communion \citep{lawson2022hiring}. 

\namedpar{Workplace proximity} and availability of reliable commutes can influence job satisfaction \citep{persson2016implicit} and candidate-employer interactions \citep{joseph2012meet}, with amplifying effects introduced by technology \citep{kayser2023linkedin}. Since discrimination has shaped residential patterns and influenced public transportation, this factor may have disproportionate impacts along racial and ethnic lines \citep{kim2016data}.

\namedpar{Wage gaps} are a key part of power differences between genders and races \citep{card2016bargaining,weichselbaumer2005meta,ec2020gender,hernandez2019bargaining}. This BCF has several concurrent causes, including expectations reinforced by the status quo and lower success in salary negotiation \citep{derenoncourt2022wealth,leibbrandt2015women,gray2019career}. Models that predict the likelihood that candidates will accept an offer can reinforce existing wage gaps.

\namedpar{\wasreplace{Centrality in social networks}{Social networks topology}} influences \wasreplace{a job seeker's}{job seekers'} awareness of openings \citep{fish2019gaps}, and their likelihood of being successfully screened by recruiters due to a successful referral \citep{jobvite2021recruiter,zhang2021unequal}.  \wasreplace{Minority candidates are less central than majority candidates in homophilic networks \citep{avin2015homophily}, leading to unequal opportunity and its entrenchment by link prediction models \wasnew{granting lower visibility to women} \citep{candela2023disentangling} \wasnew{and showing disparate performance for nodes with different degrees \citep{wang2022degree}}.}{Well-connected candidates have an inherent advantage over candidates with lower centrality; this advantage may be amplified by algorithms for link prediction \citep{wang2022degree}. The connection recommender system in a professional social network was found to underperform for women, recommending them less frequently for professional connections than men with similar success rates \citep{candela2023disentangling}. This effect may be exacerbated by homophily and biased preferential attachment \citep{avin2015homophily}.}

\subsection{Individual preferences}
\label{sec:ind_pref}

Next, we consider BCFs that are an apparent consequence of individual preferences, but represent generalized patterns for protected groups. Listing a bias under this category does not make it an individual responsibility and a reasonable ground for discrimination. On the contrary, we aim to highlight the fact that some seemingly individual choices operated by candidates actually result from wider and recurrent patterns associated with protected attributes. Therefore, employers and providers of algorithmic hiring models should carefully consider these BCFs.

\namedpar{Job satisfaction} influences job commitment \citep{bedeian1992age,shields2002racial}. Historically disadvantaged groups such as transgender, nonbinary, female, black, and disabled workers are more likely to experience discrimination and harassment on the job \citep{schneider1997job,shields2002racial,waite2021should}. This fact may be reflected in data sets as a lower tenure for these groups, which can be penalized by algorithms trained to maximize tenure in order to reduce hiring costs and retain human capital.

\namedpar{Self-promotion} gaps related to gender have been documented in the hiring domain \citep{altenburger2017are} and beyond; men tend to self-evaluate higher than women even without intrinsic incentives \citep{exley2022gender}. This reflects on the visibility and perceived competence of candidates at the different stages of the hiring pipeline; it can be especially difficult to subvert for women due to the social and economic penalties incurred for violating female gender stereotypes \citep{moss2010disruptions}. 

\namedpar{Willingness to commute} is another individual factor with gender-related patterns. Women tend to be more restrictive in their choice of job search area \citep{le2021gender,eriksson2012labor}. This difference may be related to gender roles with respect to household and childcare responsibilities. Furthermore, people with a migration background are less likely to own a motor vehicle \citep{klein2017car}.

\namedpar{Salary negotiation} differences between men and women are documented, including in propensity \citep{leibbrandt2015women} and strategy \citep{gray2019career}. Different interpretations have been advanced, including lower risk aversion and the perceived chance of success \citep{gray2019career,hernandez2019review}. Although ostensibly a personal outcome of female candidates, unsuccessful salary negotiation is also based on an unfair status quo that influences group expectations \citep{gray2019career}.

\namedpar{Culture-based avoidance} or attraction, is a self-selection BCF that influences mainly the sourcing stage. Explicitly mentioning requirements such as community outreach can signal the posture of an employer and act as a pull factor for minority job seekers \citep{smith2004interrupting,omeara2020nudging}. On the contrary, job descriptions with unrealistic requirements discourage job seekers who lack one or more requirements, with a repulsive effect that can be stronger for candidates from vulnerable groups \citep{mohr2014why,fuller2021hidden}.
Wording itself can also sustain inequality: gendered wording in job advertisements that makes use of stereotypically agentic (male) language, such as ``leadership'' or ``delivery'', can make a position less attractive for women \citep{gaucher2011evidence}. The same alienating effect can occur by signaling an unwelcoming workplace culture at other stages of the hiring pipeline \citep{wynn2018puncturing}.

\namedpar{Work gaps}, i.e. periods without formal employment, often reflect negatively on a candidate's probability of securing a job \citep{fuller2021hidden}. Gender asymmetries in caregiving responsibilities put women at a systematic disadvantage \citep{farre2016parental,lee2015more}.

\subsection{Technology blindspots}
\label{sec:tech_blind}

Finally, we describe the biases introduced by biased components integrated into larger algorithmic hiring pipelines. This non-exhaustive list aims at demonstrating the need for proactive bias-preventing reasoning also (and perhaps especially) when using off-the-shelf tools.

\namedpar{Ad delivery optimization} can skew the audiences reached by job advertisements. Multiple studies have shown that maximization of cost-effectiveness based on ad delivery metrics, such as impressions or clicks, makes delivery skewed in accordance with gender and race stereotypes in jobs \citep{datta2015automated,lambrecht2019algorithmic,angwin2017dozens,ali2019discrimination}. Ad text and images can further skew the audience \citep{ali2019discrimination}. This happens even though advertisers design neutral campaigns and is exacerbated if they target specific attributes \citep{venkatadri2020potential}, increasing the opportunity for bias and opacity at the sourcing stage. It should be noted that the platform(s) chosen to run a campaign can introduce a further bias in favor of its predominant demographics; campaigns run on platforms that cater to younger users, for example, are less likely to reach older segments of the population.

\namedpar{Accessibility issues and ableist norms} can discourage disabled people from job applications \citep{scholz2020taken} and tilt evaluations against them \citep{tilmes2022disability,trewin2019considerations}. Asynchronous video interviews can produce specific patterns from candidates with speech impairment (e.g. short answers) or mismeasure input from candidates with sight impairment (e.g. eye contact), which are judged unfavorably by algorithmic recruitment models \citep{orcaa2020description}.

\namedpar{Disparate performance of language processing and computer vision} tools has been widely demonstrated with respect to gender, race, and other sensitive attributes \citep{tatman2017gender,blodgett2016demographic,buolamwini2018gender}. Off-the-shelf algorithms from these domains integrated into hiring pipelines \citep{kaya2017multimodal} are likely to underserve minority candidates for feature extraction and negatively affect algorithms that are based on these feature.

\namedpar{Differences in platform engagement} broadly divide people into frequent and infrequent users. This causes an overrepresentation of the former in the training data, leading to rich-get-richer dynamics in job platforms \citep{nandy2022achieving}. Note that minority job seekers may suffer a lower quality of service, leading to disengagement and triggering a negative feedback loop by iteratively lowering training representation, quality of service, and engagement. Moreover, platforms are more likely to have rich profiles and metadata for their common users, while lacking information for less engaged job seekers, who are penalized due to missing data. 

\namedpar{Biased psychological assessment} performance for minorities can result in systematic disadvantages \citep{reynolds2012bias}. Systematic differences in the results of psychometric tests between subgroups \citep{hutchinson2019years,cleary1968test}, caused by low discriminant validity \citep{jacobs2021measurement} or construct contamination \citep{eignor2013standards}, can be especially problematic for tests embedded in larger resource allocation processes, such as hiring systems.

\namedpar{Background check tools} allow employers to obtain additional information on candidates from public records \citep{thompson2020countenancing}. These databases contain a wealth of pre-conviction information, such as arrest data \citep{lageson2021digitizing} whose validity as a proxy for crime is questionable \citep{fogliato2021validity}. Among other critical aspects, criminal background checks before employment run the risk of feeding the racial disparity of policing \citep{pierson2020large} and other areas of the criminal justice system \citep{kovera2019racial} into hiring systems.

\namedpar{Interaction biases} between employers and hiring technology can amplify tiny differences in algorithmic output \citep{baeza2018bias}. During the sourcing stage, for instance, recruiters tend to focus on very specific credentials, certifications, and keywords that exclude atypical individuals from the initial pool of candidates, despite having the right skills \citep{fuller2021hidden}.  More general examples include position bias that causes underexposure risks for candidates that are not at the top of a ranking \citep{craswell2008experimental} and automation bias leading to over-reliance on technology and reduced human oversight \citep{goddard2012automation,kupfer2023check}.   

\subsection{Overlaps and Interactions}
\label{sec:bcf_intersec}


In the analysis above, we identified and described how the workings of key BCFs contribute to algorithmic discrimination within a digital and non-digital hiring context. It is important to note that these factors are interlocking and overlapping insofar as BCFs are mutually reinforced by the structure of institutions (Section \ref{sec:inst_bias}), individual preferences (Section \ref{sec:ind_pref}), and technological blind spots (Section \ref{sec:tech_blind}).

Intersectional identities are particularly vulnerable as they are likely to be affected by multiple BCFs. Let us consider the case of JS, a woman with children and a migration background looking for a job in the hospitality industry. Her chances of securing suitable employment is hindered from the outset as (1) her connections to the local industries are weak and advertisements do not reach her, keeping her unaware of certain possibilities; (2) she finds out about certain openings but is discouraged from applying by a difficult commute based on public transport. Through public employment services, her data is entered into a shared database accessed by human resources companies. (3) Her profile is hastily filled in and lacks important metadata; this fact pushes her down in rankings every time her profile matches recruiters' queries, also because the time of last login influences job-seekers' rankings. (4) \wasreplace{Many job descriptions require}{Recruiters seek} a catering and hospitality degree that JS does not have, making matches with her profile infrequent. (5) Due to a lack of leadership roles in her profile, the only matches she receives are at the apprentice level and (6) she is screened out from openings employing AVI models that do not recognize her accent. She is interviewed by few businesses; most turn her down because (7) they deem her a poor cultural fit and (8) are afraid that she will apply for parental leave. One business is finally willing to hire her through an HR company. (9) Despite a mid-level payment agreement between the employer and the HR company, the latter estimates that JS will accept the minimum wage (which she does) and pockets the difference. 


This fictitious, yet plausible story sketches how multiple biases compound and reinforce each other.
However, even if intersectional discrimination occurs, there are still reasons why the right approach to technological design can reduce bias. One upshot of understanding bias as an inherently intersectional process is that it also offers a way to reduce discrimination. Since the factors that create bias are interrelated and mutually reinforcing, by halting or ameliorating one BCF, we may introduce positive feedback loops on other BCFs. 
By removing the discriminatory effect of any one factor, we can hope to reduce its influence on the other factors that reinforce each other in a discriminatory way. \wasremove{Such network effects may result in a situation where a small reduction in the bias or discrimination of algorithmic systems will compound into a large effect.} For example, an increased representation of a minority group in a company can improve job satisfaction, increase their importance and visibility, and encourage other minority applicants in future hiring rounds. With time, this contributes to reducing vertical segregation, with minority employees in key positions shaping company culture, and attracting more candidates with a similar background. This optimistic view reflects positive spillover effects \citep{matsa2011chipping,kunze2017women}.

\section{Measures}
\label{sec:measures}
Fairness measures for algorithmic hiring consider different types of systems, including AVI scoring, CV ranking, and advertising algorithms, operating at different stages and on different data modalities. In this section, we present them in a unified notation (Table \ref{tab:notation}) and discuss their key dimensions, summarized in Table \ref{tab:measures}. It is worth noting that the next three sections are based on a systematic review of the literature summarized in Appendix \ref{sec:lit_rev}.


\subsection{Dimensions of Fairness Measures}
\label{sec:fairness_dimensions}
We begin with a summary of dimensions that are important in the algorithmic fairness literature; some are established in the literature, such as \emph{flavor} \citep{mitchell2021algorithmic} and \emph{conditionality} \citep{xu2020algorithmic}, others are highlighted as important in the hiring domain, such as \emph{granularity} and \emph{interpretability}. It should be noted that we focus on group fairness; we found a single article treating individual fairness \citep{markert2022}.  

\namedpar{Flavor}. Fairness measures can target different properties of an algorithmic decision-making system. The main notions of fairness in hiring are the following.
\begin{itemize}
    \item \emph{Outcome fairness} \citep{hardt2016equality} is the most common. It looks at predictions from the perspective of candidates, measuring differences in their preferred outcome, such as being screened in, typically corresponding to a positive prediction $\hat{y}$. Systems are deemed fair from an outcome perspective if quantities such as acceptance rates ($\Pr(\hat{y}=1)$) or true positive rates ($\Pr(\hat{y}=1|y=1)$) are similar between groups. 
    \item \emph{Accuracy fairness} \citep{beutel2019fairness} takes a closer perspective on the decision maker, requesting equalization of accuracy-related properties between groups, such as the average absolute error. Measures in this family lack the notion of preferable outcomes and candidate benefits.
    \item \emph{Impact fairness} relates algorithmic outcomes to downstream benefits or harms for individuals. These measures are rare in the literature, as they require a broader understanding and modeling of the sociotechnical system around the decision-making algorithm.
    \item \emph{Process fairness} \citep{grgic2016case} is a notion that considers the equity of the procedure leading to a decision. Related to \emph{procedural justice} \citep{solum2004procedural}, process fairness has been operationalized for algorithmic decision-making based on people's approval for the use of a given feature in a specific scenario. In hiring, this has been associated with the predictability of sensitive attributes from non-sensitive ones \citep{booth2021bias}, implying that the absence of information on sensitive attributes in the data will lead to fair decision-making.
    \item \emph{Representational fairness} \citep{abbasi2019fairness} relates to stereotyping and biases in representations. In the context of algorithmic hiring, it is especially relevant for the wording of job descriptions which can skew the probability of application for different demographics \citep{gaucher2011evidence}. 
\end{itemize}

\namedpar{Conditionality}. Conditional fairness measures accept group differences, as long as they can be attributed to a set of variables deemed acceptable grounds for differentiation. This dimension is strongly connected to world views \citep{friedler2016possibility,hertweck2021moral}, and the extent to which differences in variables between sensitive groups are influenced by measurement errors and unjust social structures.

\namedpar{Granularity}. The same model can lead to different results if used in different ways, including the use of different thresholds for classifiers or different cutoffs for rankers. Measures with fine granularity take this fact into account by measuring the fairness of a system across different operating conditions, as opposed to coarse-granularity measures which consider a single point of operation.

\namedpar{Normativity}. Measures with explicit normative reasoning set a precise target for groupwise quantities, either in absolute terms or relative to each other. Implicit normative reasoning, on the other hand, is typical of measures introduced without a complete discussion of desiderata. Strong normative reasoning is, in general, preferable, but it is not a guarantee on its own. For example, while a measure providing a contextualized and accurate operationalization of a construct defined in the law displays normativity, the measure will inherit all the limitations of the underlying law.

\namedpar{Interpretability}. We consider the interpretability of group fairness along two dimensions. Direction-interpretability allows us to immediately understand whether a group is at an advantage or disadvantage by comparing the measure against a threshold. Magnitude-interpretability lets us quantify the (dis)advantage by evaluating the distance from a threshold.

\namedpar{Native Multinarity}. A measure is multinary if it accounts for more than two sensitive attribute values. Binary measures can sometimes be extended to the multinary case, whereas truly multinary measures natively account for this occurrence.

\subsection{Notation}\label{subsec:notation}

Let $x \in \mathcal{X}$ denote a vector containing non-sensitive features and let $y \in \mathcal{Y}$ be an unknown target variable, inferred at test time as $\hat{y}=f(x)$.  Furthermore, let $f_{\text{soft}}(x)$ denote a scoring function supporting estimates $\hat{y}$ via thresholding and item ranking $\tau = \text{argsort} (f_{\text{soft}}(x))$ via sorting. Furthermore, let $s \in \mathcal{S}$ indicate a vector of sensitive attributes, so that $s=g$ defines a specific protected group, and $s=\stcomp{g}$ represents its complement. Overall, a data point or item $i$ is indicated as $i=(x_i, s_i, y_i)$ and $i=\tau(k)$ denotes the item at position $k$ in ranking $\tau$. For cardinality, let $N$ denote the total number of data points in a sample of interest, let $N_g$ indicate the number of items in group $g$, and let $N_g^k$ be the number of items in $g$ among the top $k$ of a ranking $\tau$. Additionally, let $D_g \in [0,1]$ denote the desired representation for items in $g$ among the ones receiving a favorable algorithmic outcome, e.g., the top-ranked items or the positively classified ones. Finally, let $h:\mathcal{X} \rightarrow \mathcal{S} $ denote a classifier issuing predictions in $\mathcal{S}$ and let $h_{\text{soft}}(x)$ indicate its soft scoring version (e.g., issuing posterior membership probabilities).

\begin{table}[bt]
  \caption{Main notational conventions used in this work.}
  \label{tab:notation}
  \begin{center}
    \begin{tabular}{|c|p{10cm}|}
      \hline
      $x \in \mathcal{X}$ & a vector of non-sensitive attributes \\
      $s \in \mathcal{S}$ & a vector of sensitive attributes\\
      $s=g $ & a sensitive attribute value \\
      $s=\stcomp{g} $ &  complement of a sensitive attribute value \\
      $y \in \mathcal{Y}$ & target variable from domain
                            $\mathcal{Y}$\\
      $i = (x_i,s_i,y_i)$ & a data point or item \\
      $\hat{y} = f(x)$ & 
                        a classifier $f:\mathcal{X} \rightarrow \mathcal{Y}$ 
                        issuing predictions in $\mathcal{Y}$ for data points in $\mathcal{X}$ \\ 
       $f_{\text{soft}}(x)$ & a soft classifier, from which predictions $\hat{y}=f(x)$ can be derived through thresholding \\
      $\tau = \text{argsort} (f_{\text{soft}}(x))$ & ranking of items, typically sorted by target estimates\\
      $i=\tau(k)$ & item at rank $k$ in $\tau$ \\
      $N$ & number of items \\
      $N_g$ & number of items in $g$ \\
      $N_g^k$ & number of items in $g$ in the top $k$ positions of a ranking \\
      $D_g$ & desired representation for group $g$ \\
      $h(x)$ & 
                        a classifier $h:\mathcal{X} \rightarrow \mathcal{S} $ 
                        issuing predictions in $\mathcal{S}$ for data points \\ 
    $h_{\text{soft}}(x)$ & a soft classifier for sensitive attributes \\
      \hline
    \end{tabular}
  \end{center}
\end{table}

\subsection{Measures}\label{subsec:measures}

\subsubsection{Outcome Fairness} These measures summarize the perspective of candidates by focusing on their preferred outcome; they are the most common.

\textbf{Skew at $k$} ($\skewk$) \citep{geyik2019fairness} evaluates a ranking at a specific rank $k$. It computes the logarithm of the ratio between the desired representation of a sensitive group $D_g$ and the actual representation ($N_g^k / k$). 
    \begin{align}
        \skewk_g = \log \left ( \frac{N_g^k/k}{D_g} \right ) 
    \end{align}
This measure exhibits strong normativity, as it requires a precise definition of target representation $D_g$ for each sensitive group. It is direction-interpretable, as values above zero indicate an advantage for group $g$, while the presence of the $\log$ function hinders its magnitude-interpretability. As a summary for multiple groups, \citet{geyik2019fairness} consider the maximum and minimum skew values, e.g. $\text{max}$\nobreakdash-$\skewk = \maxg \skewk_g$.

\textbf{Normalized Discounted cumulative KL Divergence} ($\ndkl$)  \citep{arafan2022end,geyik2019fairness} computes a divergence between the desired distribution of the representation ($D=\{D_g, \forall g \in \mathcal{S}\}$) and the actual one at rank $k$ ($N^k=\{N_g^k/k, \forall g \in \mathcal{S}\}$). The measure consists of KL divergences, calculated at each rank, and aggregated with a logarithmic discount, making it granular. Notice that KL divergence is another way to summarize $\skewk$ for multiple groups,  weighted with a logarithmic discount. It is defined as
    \begin{align}
        \ndkl = \frac{1}{Z} \sum_{k=1}^K \frac{1}{\log_2(k+1)} d_{\text{KL}}(D, N^k) \label{eq:ndkl}
    \end{align}
 \noindent where $Z$ is a normalization constant.  Note that several factors contribute to making $\ndkl$ difficult to interpret. First, advantages and disadvantages for a group $g$ at some rank $k$ can yield the same KL divergence value. Second, advantages and disadvantages at different ranks do not compensate each other; if a group is underrepresented at low ranks, its overrepresentation at high ranks leads to even worse values of $\ndkl$, despite being a form of compensation.

\textbf{Acceptance Rate Ratio}, better known as \textbf{Disparate Impact} ($\di$) \citep{booth2021bias,burke2021fair,hemamou2021judge,kochling2021highly} is a measure of disparity in classification with a fixed threshold. Note that this measure is also used in rankings by translating a threshold (or percentile) into a cutoff rank $k$. This measure is related to US labor law, adverse impact, and the 80\% rule (Section \ref{sec:us_law}). Below, we report its popular min-over-max version, computing the selection rate ratio between the worst-off and best-off groups:
    \begin{align}
        \di &= \frac{\ming N_g^k / N_g}{\maxg N_{{g}}^k / N_{{g}}} \nonumber \\
        &= \frac{\ming \Pr(\hat{y}=1 | s=g)}{\maxg \Pr(\hat{y}=1 | s={g})} \label{eq:di}
    \end{align}
We consider this measure interpretable, so long as the best and worst-off groups are reported.

\textbf{Acceptance Rate Difference} better known as \textbf{Demographic Disparity} ($\dd$) \citep{rus2022closing}, focuses on disparities in groupwise acceptance rates, similarly to $\di$, but computes the difference instead of the ratio. We define it below for classifiers and ranking algorithms, implicitly assuming a cutoff point at rank $k$ for the latter:
    \begin{align}
        \dd &= N_g^k / N_g -  N_{\stcomp{g}}^k / N_{\stcomp{g}} \nonumber \\ 
        &= \Pr(\hat{y}=1 | s=g) - \Pr(\hat{y}=1 | s=\stcomp{g})
    \end{align}
This approach may be easier to interpret \citep{zliobaite2015survey}, but fails to capture large disparities when acceptance rates are small. If, for example, $\Pr(\hat{y}=1 | s={g})=10\mathrm{e}{-02}$ and $\Pr(\hat{y}=1 | s={\stcomp{g}})=10\mathrm{e}{-07}$, we measure a low value $\dd \simeq 10\mathrm{e}{-02}$, despite a difference of five orders of magnitude in acceptance rates.  

\textbf{Representation in Positive Predicted} ($\rpp$) \citep{ali2019discrimination,imana2021auditing,pena2020bias} is a measure used in noncooperative audits, where the overall population of interest for an algorithm is unknown. In online advertising, campaigners are informed about the size and composition of the audience reached by an ad (positive predicted class) but have no information on platform users and users who were active and thus candidates for an ad impression. This makes it impossible to compute groupwise acceptance rates ($\Pr(\hat{y}=1|s=g)$), but allows for estimates of groupwise representation in the positive predicted class ($\Pr(s=g|\hat{y}=1)$) and, trivially, its difference  with respect to the complementary group:
    \begin{align}
        \rpp_g &= \Pr(s=g|\hat{y}=1) \\
        \text{RPPD} &= \rpp_g - \rpp_{\stcomp{g}} = 2\rpp_g -1
    \end{align}

\textbf{True Positive Rate Difference ($\tprd$)} \citep{dearteaga2019bias,hemamou2022delivering} measures disparities in true positive rates (also known as \emph{recall}) between different sensitive groups. This measure is closely related to equal opportunity \citep{hardt2016equality} and separation \citep{barocas2019fairness} and presupposes the availability of a ground-truth variable $y$ to condition on:
    \begin{align}
        \tpr_g &= \Pr(\hat{y} = 1 | y=1, s=g) \nonumber \\
        \tprd &= \tpr_g - \tpr_{\stcomp{g}} 
    \end{align}

A closely related measure is the \textbf{False Negative Rate Ratio} ($\fnrr$) \citep{kochling2021highly}, defined as 
\begin{align}
        \fnr_g &= \Pr(\hat{y} = 0 | y=1, s=g) \nonumber \\
        \fnrr &= \frac{\fnr_g}{\fnr_{\stcomp{g}}} 
\end{align}

As an aggregate measure for non-binary sensitive attributes, \citet{hemamou2022delivering} propose the \textbf{Root Mean Square} ($\rms$) of the vector containing all $\tprd$ components:
\begin{align}
     \rms = \sqrt{\frac{1}{|\mathcal{S}|} \sum_{g \in \mathcal{S}} \tprd_g^2}
\end{align}

\citet{nandy2022achieving} propose an \textbf{eXtended Equality of Opportunity} ($\xeo$) measure, which is a granular version of $\tprd$. They consider a soft classifier with a variable threshold and compute the Kolmogorov-Smirnov distance between the resulting score distributions for positives in different sensitive groups.
    \begin{align}
        \xeo = \max_x [ | \Pr(f_{\text{soft}}(x) \leq t | y=1, s=g) - \Pr(f_{\text{soft}}(x) \leq t | y=1, s=\stcomp{g})|] \label{eq:xeo}
    \end{align}

\textbf{Discounted Representation Difference} ($\drd$) \citep{zhang2022are} is a granular measure of demographic disparity focused on rankings. It measures the difference in acceptance rates with a variable cutoff rank $k$ by observing the sensitive attribute of the item $\tau(k)$ and applying a logarithmic rank-based discount before updating a counter. In other words, $\drd$ measures groupwise representation as
    \begin{align}
        \drd = \sum_{k=1}^K \frac{1}{\log_2(k+1)} [\mathbbm{1}(s_{\tau(k)} = g) -  \mathbbm{1}(s_{\tau(k)} = \stcomp{g})]  
    \end{align}
This measure is interpretable since it conveys the difference in candidates' exposure across groups under a recruiter browsing model with logarithmic decay. See \citet{careterette2011system} for an introduction to browsing models.

\textbf{Score KL Divergence} ($\skl$) \citep{pena2020bias} considers the distribution of scores ($f_{\text{soft}}(x_i)$) in different groups and measures unfairness by computing their KL divergence. Let $D^f_{g}$ define the probability distribution of continuous scores $f_{\text{soft}}(x_i)$ for group $g$, then $\skl$ is defined as
    \begin{align}
        \text{SKL} = \text{KL}(D^f_{g}, D^f_{\stcomp{g}})
    \end{align}

\textbf{Log Rank Regression} ($\lrr$) \citep{chen2018investigating} is a measure defined for search engines by fitting a linear model on the logarithm of rank $k=\tau^{-1}(i)$ at which candidates are presented on the result page. Independent variables comprise both sensitive $s$ and non-sensitive attributes $x$, where the latter include skills and education. $\lrr$ reports the coefficient (and $p$ value) associated with the sensitive attribute. Fitting models to quantify the influence of a sensitive feature on an outcome variable is typical of situations with limited access to the model(s) responsible for the outcomes. This is a common setting for external noncooperative audits.
    \begin{align}
        \log (\tau^{-1}(i)) &= \beta_x x_i + \beta_s s_i + \mu + \epsilon \nonumber \\
        \lrr &= \hat{\beta}_s \label{eq:lrr}
    \end{align}
A similar approach is proposed in \citet{lambrecht2019algorithmic} to estimate the influence of sensitive attributes such as age and gender in advertisement delivery optimization. This measure considers differences in outcomes acceptable, so long as they can be explained by non-sensitive attributes $x$.

\textbf{Mean Error Difference} ($\med$) \citep{singhania2020grading}  focuses on regression, measuring systematic groupwise biases. Following a more general paradigm, $\med$ caters to the multinary case by considering the maximum and minimum (signed) bias.
    \begin{align}
        \med = \maxg \frac{1}{N_g} \sum_{i \in g} (y_i-f_{\text{soft}}(x_i)) - \ming \frac{1}{N_g} \sum_{i \in g} (y_i-f_{\text{soft}}(x_i)) \label{eq:med}
    \end{align}

\subsubsection{Accuracy fairness} These measures study model accuracy across sensitive groups, aligning more closely with the perspective of decision makers. They are all conditional on the target variable $y$, inheriting the biases encoded in the ground truth.

\textbf{Mean Absolute Error} ($\mae$) \citep{yan2020mitigating,singhania2020grading} targets regression problems similarly to $\med$ (Eq.~\ref{eq:med}). It compares groupwise accuracy by measuring the absolute error for each individual and computing the average for items in the same sensitive group.
    \begin{align}
        \text{MAE} = \frac{1}{N_g} \sum_{i \in g} |y_i-f_{\text{soft}}(x_i)| - \frac{1}{N_{\stcomp{g}}} \sum_{i \in \stcomp{g}} |y_i-f_{\text{soft}}(x_i)|
    \end{align}

\textbf{Balanced Classification Rate Difference} ($\bcrd$) \citep{kochling2021highly} is a measure of the disparity in classification accuracy between groups. It targets the balanced classification rate, defined as the average between the true positive and the true negative rate:
    \begin{align}
        \text{BCR}_g &= \frac{\tpr_g + \tnr_g}{2} \nonumber \\
        \bcrd &= \text{BCR}_g - \text{BCR}_{\stcomp{g}}
    \end{align}

\textbf{Mutual Information Difference} ($\midif$) \citep{kochling2021highly} is another measure of classification accuracy fairness. It computes model accuracy within a group $g$ as the mutual information between the target $y$ and the prediction $\hat{y}$ for the items in $g$ and compares it against the mutual information for the remaining items. 
    \begin{align}
        \text{MI}_g &= \text{MI}_{\{i \in g\}}(\hat{y}_i,y_i) \nonumber \\
        \midif &= \text{MI}_g - \text{MI}_{\stcomp{g}}
    \end{align}

\begin{table}[t]
    \caption{Main measures of fair hiring and their properties. Columns report the fairness dimensions described in Section \ref{sec:fairness_dimensions}.}
    \label{tab:measures}
     \begin{threeparttable}
        \centering
        \begin{tabular}{|r||p {2cm}|c|c|c|c|c|c|}
          \hline
          & \textbf{used in} & \textbf{flavor}  & \textbf{cond.}  & \textbf{multi.}& \textbf{gran.} & \textbf{norm.} & \textbf{interp.}  \\
          \hline\hline
           $\skewk$ & \citep{geyik2019fairness} & outcome & & \no & \no & \checkmark & D \\ \hline
           $\ndkl$ & \citep{geyik2019fairness,arafan2022end} & outcome & & \checkmark & \checkmark & \checkmark & \no \\ \hline
           $\di$ & \citep{booth2021bias,burke2021fair,kochling2021highly,kochling2021highly,wilson2021building,deshpande2020mitigating} & outcome &  & \checkmark & \no  & \checkmark & \checkmark \\ \hline
           $\dd$ & \citep{rus2022closing} & outcome &  & \no & \no  & \checkmark & \checkmark \\ \hline
           $\rpp$ & \citep{ali2019discrimination,imana2021auditing,pena2020bias} & outcome &  & \no & \no  & \no & \checkmark \\ \hline
           $\drd$ & \citep{zhang2022are} & outcome &  & \no & \checkmark & \checkmark & \checkmark \\ \hline
           $\skl$ & \citep{pena2020bias} & outcome &  & \no & \checkmark & \checkmark & \no \\ \hline
           $\lrr$ & \citep{chen2018investigating} & outcome & $x$ & \no  & \checkmark  & \no & D \\ \hline
           $\tprd$ & \citep{dearteaga2019bias,hemamou2022delivering} & outcome & $y$ & \no & \no & \checkmark & \checkmark \\ \hline
           $\fnrr$ & \citep{kochling2021highly} & outcome & $y$ & \no & \no & \checkmark & \checkmark \\ \hline
           $\med$ & \citep{singhania2020grading} & outcome & $y$  & \checkmark & \checkmark & \checkmark & \checkmark \\ \hline
           $\rms$ & \citep{hemamou2022delivering} & outcome & $y$ & \checkmark & \no & \no & \no \\ \hline
           $\xeo$ & \citep{nandy2022achieving} & outcome & $y$ & \no & \checkmark & \no & \no \\ \hline
           $\mae$ & \citep{yan2020mitigating,singhania2020grading} & accuracy & $y$  & \no & \checkmark & \checkmark & \no \\ \hline
            $\bcrd$ & \citep{kochling2021highly} & accuracy & $y$ & \no & \no & \checkmark & \no \\ \hline
            $\midif$ & \citep{kochling2021highly} & accuracy & $y$ & \no & \no & \checkmark & \no \\ \hline
            $\sd$ & \citep{rus2022closing} & impact &  & \no & \no & \checkmark & \checkmark \\ \hline
            $\gbs$ & \citep{hu2022balancing} & representational &  & \no & n.a & \no & \checkmark \\ \hline
            $\sauc$ & \citep{parasurama2022gendered2,parasurama2021degendering,booth2021bias,hemamou2021judge,rus2022closing,hemamou2022delivering} & process &  & \no & n.a. & \no & n.a. \\ \hline
            $\gtr$ & \citep{parasurama2022gendered2} & process & $x$ & \no & n.a. & \no & n.a. \\ \hline
            $\mia$ & \citep{yan2020mitigating} & process & $y$ & \checkmark & \no & \checkmark & n.a. \\ \hline
        \end{tabular}
        \begin{tablenotes}
            \tiny
            \item * In the last column, ``D'' denotes direction-interpretability, for measures which clearly convey whether one group is at an advantage; \checkmark indicates both direction- and magnitude-interpretability, the latter being assigned to measures intuitively quantifying the advantage; ``\no'' indicates none of the above; ``n.a.'' in process and representational fairness stand for \emph{not applicable}.
        \end{tablenotes}
     \end{threeparttable}
    \end{table}

\subsubsection{Impact fairness} This flavor of fairness models the impact of algorithmic outcomes on data subjects, measuring differences in harms and benefits between populations.

\textbf{Salary Difference} ($\sd$) \citep{rus2022closing} is the only measure of this family proposed in the literature. Considering an embedding-based model, each candidate is matched to the closest vacancy in the embedding space. Based on the average wage for the role described in the vacancy, an average salary is calculated for male and female candidates in the same industry, and their difference quantifies the gender impact of the model on earnings in that industry.
\begin{align}
    f(x_i) &: \text{closest job match for candidate $i$} \nonumber \\
    W(f(x_i)) &: \text{average wage for job $f(x_i)$} \nonumber \\
    \sd &= \frac{1}{N_g}\sum_{i \in g} W(f(x_i)) - \frac{1}{N_{\stcomp{g}}}\sum_{i \in \stcomp{g}} W(f(x_i))
\end{align}
    


\subsubsection{Representational fairness} Measures to quantify stereotypes and representational harms are relatively understudied \citep{abbasi2019fairness,fabris2020gender}, despite being a fundamental driver in the reinforcement of societal biases.

\textbf{Gender Bias Score} ($\gbs$) \citep{hu2022balancing} is a coarse measure of bias toward ``masculine'' or ``feminine language'' in job descriptions. It takes advantage of the word inventory from \citet{konnikov2022bias}, which comprises traits associated with gender roles in labor, so that each term $t$ in a job description can theoretically be coded as belonging to a set of stereotypically masculine, feminine, or neutral traits. The measure is defined as
\begin{align}
    \gbs = \text{sign}(x_m-x_f) \cdot \max \left \lbrace \frac{x_m-x_f}{x_m}, \frac{x_f-x_m}{x_f} \right \rbrace,
\end{align}
where $x_m$ ($x_f$) represents the number of stereotypically male (female) words in a job description.

\subsubsection{Process fairness} These notions of fairness operationalize desiderata about decision-making beyond outcomes. Frequently,  they derive from judgments of admissibility for specific variables $x$ in a given scenario. In algorithmic hiring, the literature focuses on the amount of information on sensitive attributes $s$ encoded by non-sensitive features $x$ and the target variable $y$, with the goal of minimizing it.   

\textbf{Sensitive AUC} ($\sauc$) \citep{parasurama2022gendered2,parasurama2021degendering,booth2021bias,hemamou2021judge,rus2022closing,hemamou2022delivering} is a widely used measure of the information about sensitive attributes stored in (proxy) non-sensitive attributes. It is based on training a classifier $h(x)$ for the sensitive attribute $s$ on non-sensitive features $x$; its accuracy is evaluated as the Area Under the receiver operating characteristic Curve (AUC). If little information on sensitive attributes can be recovered from the variables employed in the decisions, then the decision is deemed procedurally fair according to this definition.
\begin{align}
    \sauc = \text{AUC}(h_{\text{soft}}(x))
\end{align}
\citet{hemamou2021judge} extend ($\sauc$) to the multi-class case by training one-vs-all classifiers and reporting their maximum AUC value.

\textbf{Ground Truth Regression} ($\gtr$) \citep{parasurama2022gendered2} regresses $(x,s)$ on $y$, and focuses on the latter coefficient ($\hat{\beta}_s$) similarly to $\lrr$ (Eq.~\ref{eq:lrr}), with the key difference that the target of the regression is the target variable. This measure estimates the importance of the sensitive attribute to predict the target, conditional on non-sensitive attributes. In other words, $\gtr$ audits the biases encoded in the target variable, recognizing it as a key factor for the decision-making process, regardless of the predictive model employed:
    \begin{align}
        \log \left ( \frac{y_i}{1-y_i} \right ) &= \beta_x x_i + \beta_s s_i + \mu + \epsilon \nonumber \\
        \gtr &= \beta_s
    \end{align}
Both $\sauc$ and $\gtr$ are \emph{ex-ante} measures, i.e. they can be computed from the data before the final model is trained; therefore, the \emph{granularity} dimension does not apply to them.

\textbf{Mutual Information Amplification} ($\mia$) \citep{yan2020mitigating} is another measure of process fairness, which considers the process fair if it does not reveal more information about sensitive attributes than the ground truth does. This metric computes the mutual information between first $\hat{y}$ and $s$, and second between $y$ and  $s$; positive differences between the two are considered an undesirable amplification of information about sensitive attributes leaked through predictions. 
    \begin{align}
        \mia = \text{MI}(\hat{y},s) - \text{MI}(y,s)
    \end{align}
Differently from $\sauc$ and $\gtr$, $\mia$ is an \emph{ex-post}, model-dependent measure.




\subsection{Discussion}

\new{In this section, we describe the main characteristics of the surveyed measures and how they inform fairness measurement choices.}

\namedpar{Choosing a measure}.
\new{We introduce a made-up scenario to exemplify deliberation around fairness monitoring. 
\emph{EquiHire}, a fictitious HR company, is choosing a fairness measure to monitor its candidate search systems for fairness with respect to ethnicity. For the sake of simplicity, they restrict the realm of possibilities to Table \ref{tab:measures}.  Since ethnicity is a multinary sensitive attribute, they discard binary measures such as $\skewk$ and $\dd$. Moreover, EquiHire practitioners want to describe their equity strategy in white papers and external communication. Therefore, they prioritize an interpretable measure, excluding complex measures such as $\ndkl$ and $\skl$ which would be difficult to explain. Although they have access to signals related to candidate fitness ($y$), they become available with great delay after hiring decisions are taken. To assess system fairness in a timely manner, they discard $y$-conditional measures such as $\tpr$ and $\rms$. Finally, since they have visibility into the downstream hiring decisions, they can use a low-granularity measure. These considerations lead EquiHire to choose $\di$ as their primary fairness measure. For a complementary angle focused on outcome fairness, they choose to monitor wage differences for employees hired through their systems, adopting $\sd$ as an additional measure. 
}

\new{
To reiterate, this is just one fictitious scenario and not a fixed recommendation for choosing the right measure. It demonstrates the importance of key characteristics of the measures summarized in Table \ref{tab:measures} for practitioner decisions; in the paragraphs below, we expand on the factors driving these decisions. 
}

\namedpar{Fairness diagnostics vs. fairness optimization}. 
Ease of interpretation is a desirable property, especially for fairness measures, as confirmed by the popularity of DI. Nevertheless, several \wasreplace{measures}{fairness measures adopted in hiring} are difficult to read in absolute terms, making it \wasreplace{difficult to connect their values to model properties}{challenging to understand the severity of fairness violations based on their values}. These values can only be interpreted relative to one another, i.e. it is clear that one value is more desirable than another, making them a possible target for optimization but less suited for diagnostics. \new{This is especially true for measures of accuracy fairness such as $\bcrd$ and $\midif$.}  Measures of outcome and impact fairness\new{, such as $\dd$ and $\sd$,} tend to be more interpretable, since they are typically defined as differences between the probability of obtaining desirable outcomes for different groups. This holds especially for binary measures, while multinary measures, such as $\rms$, often sacrifice interpretability to summarize disparities across multiple groups.

\namedpar{\wasreplace{Flexibility of use and granularity of evaluation}{Blurred lines between ranking and classification}}. In the literature, classification and ranking are typically considered separate tasks with separate algorithmic fairness measures. In hiring, we find ranking systems evaluated according to classification measures such as $\di$ \wasnew{\citep{burke2021fair}}, \wasreplace{classifiers evaluated at multiple cutoffs}{selection tasks solved with rankers \citep{burke2021fair}}, and target variables of similar nature encoded as binary, multinary, or continuous (Section \ref{sec:data}). For this reason, the separation between classification and ranking is less rigid than in other domains, and the question of which measures are best suited to evaluate these systems arises. A fundamental part of the answer lies in the expected flexibility of use for these models. If the key aspects of their usage are fixed, including the cutoff thresholds for different outcomes, then a measure with coarse granularity, considering a single operating condition (typical of the fair classification literature\new{ -- e.g. $\dd$, $\tprd$}), is most suited for evaluation. On the contrary, for models whose operating conditions are more uncertain, requiring human interaction or threshold fine-tuning, it is preferable to adopt a measure with finer granularity (often derived from the fair ranking literature\new{ -- e.g. $\ndkl$, $\drd$}), \replace{e.g. averaging}{which averages} outcomes across multiple realizations of a browsing model.

\namedpar{Diagnosing biases}.
\new{Fairness measures are especially useful when they provide actionable diagnostics by highlighting specific biases described in Section \ref{sec:bcf}. Measures of process fairness such as $\sauc$ and $\mia$ detect the presence of sensitive attribute proxies. Dedicated proxy reduction approaches (Section \ref{sec:miti}) can mitigate the resulting risk of discrimination. Representational fairness ($\gbs$) captures biases in representations; when applied to job descriptions, extreme values highlight stereotypically gendered language, signaling a risk of culture-based avoidance. Impact fairness can highlight nuanced aspects of hiring beyond selection rates; $\sd$, for instance, captures wage gaps and salary negotiation differences prompting further analysis into remuneration packages.  Measures of outcome (e.g. $\di$) and accuracy fairness (e.g. $\mae$) can serve as high-level metrics by highlighting the effects of multiple factors acting together. To exemplify, a combination of job segregation, work gaps, and disparate performance of language processing tools, even if relatively weak on their own, may trigger a very low (unfair) value of DI when they compound.}

\namedpar{\wasreplace{Abstraction yields limited measures}{Assorted fairness flavors}}.  \new{No single measure provides a complete picture. Different fairness flavors have complementary strengths and weaknesses.} Most of the fairness measures considered in hiring focus on outcome equity; the field is strongly influenced by the 80\% rule (Section \ref{sec:us_law}), which is appreciated as a quantitative rule of thumb, but also often criticized \citep{wax2011disparate,eeoc2023select}. 
Indeed, outcome fairness is based on a narrow view of algorithmic hiring as a single decision point abstracted away from its context, where equity is purely a function of algorithmic estimates and (sometimes) their similarity to target variables approximating a ground truth. Similar criticism can be moved to accuracy fairness, with the additional limitation that it is less interpretable and less aligned with the desiderata of job seekers. 
Departing from this approach, process fairness is operationalized with reference to the information on sensitive attributes encoded in the remaining variables. More work is required to understand process fairness in hiring more broadly and to suitably (if at all) operationalize it quantitatively. 
Impact fairness quantifies the downstream impacts of algorithmic outcomes on different populations; modeling important aspects of the surrounding socio-technical system, it goes beyond brittle algorithmic abstraction. Since these measures are exceedingly rare, we highlight the need for more context-specific research to understand and model the benefits and harms of decisions on job candidates at different stages of the hiring pipeline. \new{Overall, practitioners developing equity monitoring protocols should consider multiple fairness angles to gain a nuanced and more complete picture.}

\namedpar{Ignoring privileged and disadvantaged groups}. The definition of sensitive attributes in anti-discrimination law is informed by the historical (dis)advantage of specific groups \citep{simson2024lazy}. Gender and race, for example, are considered sensitive attributes due to the recurrent structural disadvantages incurred by women and black people. This is especially true in hiring, where surveys and meta-analyses find consistent biases against women and ethnic minorities \citep{bertrand2004emily,kline2022systemic,quillian2017meta}. Although it is commonly held, especially in fairness research, that disparities against disadvantaged groups should be mitigated, there is less consensus on how to treat algorithms that happen to favor historically disadvantaged communities. In other words, should algorithmic fairness be symmetrical and reject a priori notions of privileged and disadvantaged groups? This is an important normative question, \wasnew{seldom acknowledged in the fairness literature,} with immediate consequences on the choice of \wasreplace{a fairness measure}{measures}. For example, the version of $\di$ reported in Equation \eqref{eq:di} computes the ratio between the acceptance rates of the worse-off group ($\min$) and the best-off group ($\max$), 
ignoring prior knowledge of structural inequality and affected communities. 

\section{Mitigation Strategies}
\label{sec:miti}

Several algorithms have been proposed in the literature to improve model fairness; they are summarized in Table \ref{tab:miti}. We present them distinguishing between pre-processing, in-processing, and post-processing algorithms, depending on their applicability before, during, and after training.

\subsection{Pre-processing}

 \textbf{Rule-based} approaches perform a set of manipulations, typically defined by experts, on text data. They are related to process fairness (Section \ref{sec:measures}), as they focus on reducing the amount of information on sensitive attributes contained in non-sensitive features, i.e. to remove sensitive attribute proxies (Section \ref{sec:bcf}). \textbf{Rule-based Scraping} \citep{dearteaga2019bias} is a heuristic  \wasnew{for \textbf{text data}} focused on gender, aimed at removing all words that explicitly refer to the gender of a person, including first names and titles. \wasnew{In general, it acts as a feature transformation $x' = \text{scrp}(x)$. Under bag-of-words representations, the scraping function is a feature selection mechanism that removes features in a censored vocabulary~$\mathcal{V}$.
\begin{align*}
     \text{scrp} &: \mathcal{X} \rightarrow \mathcal{X}' \subseteq \mathcal{X} \\
     \text{scrp}(x^j)&= 
\begin{cases}
    x^j,& \text{if } x^j \notin \mathcal{V}\\
    \emptyset,              & \text{if } x^j \in \mathcal{V}
\end{cases}
 \end{align*}}
\noindent \citet{dearteaga2019bias} study the effectiveness of this approach in \wasnew{\textbf{job classification}} from short biographies. They find it to be moderately effective in reducing $\tprd$. \citet{parasurama2021degendering} take this approach one step further on a CV screening application. They define several rules for removing other strings related to gender, including email addresses, intrinsically gendered words (e.g. ``waitress''), and hobbies. \textbf{Rule-based Substitution} \citep{rus2022closing} is a closely related approach, which replaces intrinsically gendered words with neutral words; for example, ``his'' is changed into ``theirs''. Due to redundant encoding of sensitive information in the data, rule-based approaches often represent a weak baseline with limited impact on fairness. 

\textbf{Importance-based Scraping} is an adversarial approach for feature removal based on a sensitive attribute classifier. \citet{parasurama2021degendering} consider a \wasnew{\textbf{resume screening}} system and exploit contextualized word representations to train a gender classifier, and iteratively scrape the words with the largest feature importance for gender classification. \wasnew{In other words, they train a sensitive attribute classifier 
\begin{align*}
 \hat{s}&=h(x)   
\end{align*}
and assess the importance of features by selectively removing them from the prediction task. Proxy features $\mathcal{P}$ found to be most predictive of $s$ (highest marginal contribution to $h(\cdot)$, e.g. measured via SHAP) are scraped:
\begin{align*}
     \text{scrp}(x^j)&= 
\begin{cases}
    x^j,& \text{if } x^j \notin \mathcal{P}\\
    \emptyset,              & \text{if } x^j \in \mathcal{P}
\end{cases}
\end{align*}}
\wasreplace{They}{\citet{parasurama2021degendering}} show that SHAP-based scraping can achieve a sizeable $\sauc$ drop with a limited negative impact on performance. \citet{booth2021bias} develop an equivalent scheme for iterative feature removal in \wasnew{\textbf{asynchronous video interview}} analysis from multimodal data. They confirm the suitability of this method to improve process fairness ($\sauc$); outcome fairness ($\di$) also improves, but only for systems that were initially very unfair.

\textbf{Subspace Projection} is a common approach to reduce gender bias in \wasnew{\textbf{text}} representations based on word embeddings. Initially proposed by \citet{bolukbasi2016man}, this method is based on the observation that most intrinsically gendered information in word embeddings, such as the difference between the vectors for ``mother'' and ``father'', is contained within a small gender subspace. The algorithm is based on re-embedding each word by projecting it orthogonally to this space. \wasnew{Let $\vec{w}$ denote the embedding of a word and $G$ the gender subspace. Each word is re-embedded to
\begin{align*}
    \vec{w}' &= \frac{\vec{w}-\text{proj}_G\vec{w}}{||\vec{w}-\text{proj}_G\vec{w}||}
\end{align*}}
In the algorithmic hiring literature, subspace projection is used in \citet{parasurama2021degendering} to reduce gender biases in a \wasnew{\textbf{resume screening}} application. Although in theory this approach is suitable to remove gender proxies and contrast the negative effects of stereotype violations, it proves ineffective in reducing the amount of gender information contained in resumes, as measured by $\sauc$, in alignment with prior art \citep{gonen2019lipstick}.

\textbf{Balanced Sampling} \wasreplace{reduces}{is  a broadly applicable scheme to reduce} correlations between sensitive attributes and target variables in the training set. \citet{arafan2022end} propose a down-sampling scheme to achieve an equal representation of sensitive groups among positive and negative points in the training set. For a binary sensitive attribute, this condition can be formulated as
\begin{align*}
    \Pr_{\sigma}(s_i = g | y_i = \overline{y}) = \Pr_{\sigma}(s_i = \stcomp{g} | y_i = \overline{y}) \text{, } \forall \overline{y} \in \mathcal{Y},
\end{align*}
\noindent where we let $\sigma$ denote an algorithm's training set. \citet{yan2020mitigating} propose an upsampling approach to enforce the same condition before training a multimodal data fusion algorithm for the analysis of \wasnew{\textbf{asynchronous video interviews}} \citep{kaya2017multimodal}. Overall, balanced sampling can reduce the influence of job segregation on hiring algorithms. 

\textbf{Group Norming} is a feature manipulation approach \wasreplace{aimed at}{for \textbf{tabular data}} enforcing a similar feature distribution in all sensitive groups, through normalization. \citet{booth2021bias} propose \textbf{Groupwise z-Normalization}, i.e. they divide data points based on their sensitive group membership, and standardize each feature by subtracting the groupwise mean and dividing by the groupwise standard deviation.
\begin{align*}
    \mu_g &= \frac{1}{N_g} \sum_{i \in g} x_i \\
    \text{std}_g &= \frac{1}{N_g} \sum_{i \in g} (x_i - \mu_g)^2 \\
    x_i'&= \frac{x_i-\mu_g}{\text{std}_g} \text{ if } i \in g 
\end{align*}
This approach represents an intermediate view on conditional discrimination: it prohibits \emph{inter-group} discrimination based on a specific feature, while allowing it as a basis for \emph{intra-group} discrimination. It is tested on multimodal data and found to decrease gender predictability ($\sauc$) for paraverbal and visual data, while increasing predictability for verbal features and only marginally improving $\di$. It is unclear how to generalize this method under multiple protected attributes (e.g. race and gender); a separate application to each intersectional group (e.g. black women) is the most straightforward extension, but increasingly smaller groups run the risk of unstable normalization. Considering its limited impact on both process and outcome fairness, in conjunction with its controversy in US employment law \citep{levin2018gender,bent2019algorithmic},  the use of Group Norming is not recommended.

\begin{table}[t]
\scriptsize
  \caption{Bias mitigation methods to improve fairness in hiring}
  \label{tab:miti}
  \begin{threeparttable}
  \centering
    \begin{tabular}{|p{2.0 cm}||x {0.5 cm}|c|x {1.4 cm}|x {0.8 cm} |x {0.7 cm}|x{1.1cm}|p {3.5cm}| }
      \hline
      & \textbf{in} & \textbf{family} & \textbf{measures} & \new{\textbf{$s$ availability}} & \textbf{mode} & \textbf{approach} & \textbf{summary}  \\
      \hline\hline
       Rule-based Scraping or Substitution & \citep{dearteaga2019bias,parasurama2021degendering,rus2022closing} & PRE & $\sauc$, $\dd$, $\tprd$, $\sd$ & \new{none} & text & proxy reduction & remove or substitute gender identifiers based on hard-coded rules \\ \hline
       Importance-based Scraping & \citep{parasurama2021degendering,booth2021bias} & PRE & $\sauc$, $\di$ & \new{train}  &  & proxy reduction & iteratively remove proxy features most predictive of sensitive attribute \\ \hline
       Balanced Sampling & \citep{arafan2022end,yan2020mitigating} & PRE & $\ndkl$, $\mia$, $\mae$ & \new{train}  & & rebalancing & resample training set with same group cardinality in each class  \\ \hline
       Group Norming & \citep{booth2021bias} & PRE & $\sauc$, $\di$ & \new{train}  &  & groupwise feature transform & $z$-score normalization of each feature within each group  \\ \hline
       Subspace Projection & \citep{parasurama2021degendering} & PRE & $\sauc$ & \new{none}  & text & proxy reduction & remove gender information via embedding projection \\ \hline
        Adversarial Inference & \citep{hemamou2021judge,rus2022closing,yan2020mitigating,pena2020bias} & IN & $\sauc$, $\dd$, $\tprd$, $\sd$, $\mia$, $\mae$, $\skl$ & \new{train}  &  & proxy reduction & reduce sensitive info in latent representation through additional adv. loss   \\ \hline
        Face Decorrelation & \citep{hemamou2021judge} & IN & $\sauc$, $\di$ & \new{none}  & image & proxy reduction & discourage intermediate representations correlated with face ID through adv. loss   \\ \hline
        Name Decorrelation & \citep{hemamou2022delivering} & IN & $\sauc$, $\tprd$, $\rms$ & \new{none}  & text  & proxy reduction & discourage intermediate representation correlated with names by minimizing MI   \\ \hline
        Fair TF-IDF & \citep{deshpande2020mitigating} & IN & $\di$ & \new{train}  & text  & proxy reduction & reduce feature weight according to its group-specificity \\ \hline
        DetGreedy & \citep{geyik2019fairness,arafan2022end,suhr2021does} & POST & $\ndkl$ & \new{runtime}  & & output re-ranking & re-rank items enforcing desired group representation in each prefix \\ \hline
       CDF Rescoring & \citep{nandy2022achieving} & POST & $\xeo$ & \new{runtime}  & & output re-rescoring & re-score items with groupwise CDF trick \\ \hline
        Spatial Partitioning & \citep{burke2021fair} & POST & $\di$ & \new{train}  &  & output re-ranking & promote candidates based on group membership probability  \\ \hline
    \end{tabular}
  \end{threeparttable}
\end{table}

\subsection{In-processing}

\textbf{Adversarial Inference} \citep{rus2022closing,hemamou2021judge,yan2020mitigating,pena2020bias} is an in-processing approach to remove sensitive information from latent representations \wasnew{that can be applied across tasks (e.g. classification, ranking) and data modalities (e.g. tabular, images).} It leverages process fairness to reduce sensitive attribute proxies in pursuit of outcome fairness \citep{edwards2016censoring}. This approach directly models an adversary trying to infer an individual's sensitive attributes $s_i$ from their latent representation $l(x_i)$. The latent representation is typically derived in one or more layers of a neural architecture whose goal is to predict the target variable $y$ associated with employability. The adversarial loss for inference is defined as
\begin{align*}
    L^{\text{ADV}}_{\text{INF}} = \frac{1}{N} \sum_i \text{dist}(s_i, d(l(x_i)))
\end{align*}
\noindent where $d(\cdot)$ represents the layer(s) in the adversarial branch, so that $h(x)=d(l(x_i))=\hat{s}_i$ is the inferred sensitive attribute value, and $\text{dist}(\cdot)$ computes its distance from the actual value $s_i$. \citet{rus2022closing} show the suitability of this method to improve selected process, outcome, and impact fairness indicators in a \wasnew{\textbf{job recommendation}} scenario, while \citet{pena2020bias} demonstrate $\skl$ improvements in a multimodal setting based on synthetic \wasnew{\textbf{resume ranking}}. 

\textbf{Face Decorrelation} \citep{hemamou2021judge} was proposed for asynchronous video interview systems and, more generally, algorithms that process \wasnew{\textbf{face images}.} This approach is based on the key assumption that face data contain enough information on sensitive features such as gender and race, so that successful debiasing can be achieved without explicit sensitive attribute information. 
More in detail, let $l(x)$ denote a latent representation from a neural architecture used for employability prediction, and let $w(x)$ denote a representation of the candidates' face obtained from a state-of-the-art feature extraction method for face recognition \citep{deng2019arcface}. \citet{hemamou2021judge} propose two schemes based on Mean Squared Error (MSE) and Negative Sampling (NS). Under MSE, the adversary branch tries to leverage latent representations to reconstruct face features by minimizing
\begin{align*}
    L^{\text{ADV}}_{\text{MSE}} = \frac{1}{N} \sum_i [d(l(x_i)) - w(x_i)]^2
\end{align*}
\noindent where $d(\cdot)$ represents the final dense layer in the adversarial branch. Under NS, the adversary exploits the latent representation $l(x_i)$ extracted from an interview to discriminate the respective candidate from the remaining ones by maximizing the softmax
\begin{align*}
    L^{\text{ADV}}_{\text{NS}}(x_i) = \frac{\text{exp}(\text{sim}(l(x_i), w(x_i)))}{\sum_{j \neq i} \text{exp}(\text{sim}(l(x_i), w(x_j)))}
\end{align*}
\noindent where $\text{sim}(\cdot)$ represents a similarity function between the face recognition features $w(x_i)$ and the representations $l(x_i)$ learnt by the main branch of the network. Both variants of Adversarial Removal (NS, MSE) are shown reduce the sensitive information encoded in latent representations: notably, their $\sauc$ is on par with adversarial methods explicitly trained to predict gender and ethnicity, and may be suited for settings where sensitive attributes are unavailable during training. The effectiveness of Face Decorrelation for outcome fairness is more nuanced, showing positive effects on $\di$ for very unfair models based on the video modality and limited effects on more equitable systems based on language and audio.

\textbf{Name Decorrelation} \citep{hemamou2022delivering} is a similar approach proposed for \wasnew{\textbf{text}} data, based on sensitive information encoded in \wasnew{\textbf{names}.} The goal of this method is to reduce the Mutual Information (MI) between the representations of individuals, such as document embeddings extracted from their resumes, and a word embedding of their name. To handle the complexity of MI estimation in high-dimensional continuous spaces, the method focuses on the MI between the latent representation of the input $l(x_i)$ and a low-dimensional projection of their name $\tilde{t}_i$. The adversarial loss function is defined as
\begin{align*}
    L^{\text{ADV}}_{\text{name}} = \frac{1}{N} \sum_i \hat{\text{MI}}(\tilde{t}_i,l(x_i))
\end{align*}
\noindent 
This approach reduces the ability of adversaries to infer an individual's gender and ethnicity from their hidden representations as measured by $\sauc$\wasnew{ in a \textbf{job classification} task.} In terms of outcome fairness,  Name Decorrelation is found to improve $\tprd$ and $\rms$ with respect to gender but not ethnicity. It is worth noting that the original disparities achieved by a vanilla model, measured by $\rms$ and $\tprd$, were smaller for ethnicity than for gender. Similarly to \citet{hemamou2021judge}, this result suggests that targeting this type of process fairness can improve outcome fairness in highly imbalanced situations, while only providing limited benefits in situations with lower inequity.

\textbf{Feature Weighting} schemes learn to decrease the importance of a feature based on its likelihood of increasing the unfairness of a model. \textbf{Fair TF-IDF} \citep{deshpande2020mitigating} is an in-processing approach for \wasnew{\textbf{text}} classification and ranking applications, such as \wasnew{\textbf{resume filtering}.} As the name suggests, this algorithm is an extension of TF-IDF \citep{sparck1972statistical}, one of the most popular and influential algorithms in text search engines. In response to a query describing a job, TF-IDF produces a score $f_{\text{soft}}(x_i)$ for each resume based on the count of query terms in item $i$ and on the specificity of the terms in the entire resume collection. In other words, let $t$ denote a term (word), potentially present in a resume $i$ and in a query $q$ describing a job posting; $\text{TF-IDF}(i,t)$ is the product of a term frequency $\text{tf}(i,t)$ and an inverse document frequency $\text{idf}(t)$. The $\text{tf}(i,t)$ factor summarizes the importance of $t$ in resume $i$ by counting the occurrences of $t$ in $i$. The $\text{idf}(t)$ factor conveys the specificity of term $t$ by counting how many resumes in the entire collection contain the term $t$ and defining $\text{idf}(t)$ as its inverse; $\text{idf}(t)$ can be seen as a weight that increases (decreases) the importance of rare (common) words. \citet{deshpande2020mitigating} propose a further weighting scheme based on penalizing terms that are highly specific to a given group. 
\begin{align*}
    p\text{-ratio}(t) &= \frac{\ming{\Pr(t \in i | i \in g)}}{\maxg{\Pr(t \in i | i \in g)}} \\
    \text{fair-tf-idf}(i,t) &= \text{tf(i,t)} \cdot \text{idf(t)} \cdot p\text{-ratio(t)} \\
    f_{\text{soft}}(i,q) &= \sum_{t \in q} \text{fair-tf-idf}(i,t)
\end{align*}
\noindent The authors also propose more complex weighting schemes for the fairness factor $p\text{-ratio}(t)$, and test their ability to improve system fairness as measured by $\di$.

\subsection{Post-processing}

\textbf{DetGreedy} \citep{geyik2019fairness} and its variants are post-processing methods for \wasnew{\textbf{ranking}} algorithms targeting $\ndkl$. Given a desired representation of sensitive groups, expressed as a target distribution $D=\{D_g, \forall g \in \mathcal{S}\}$ (Eq.~\ref{eq:ndkl}), \citet{geyik2019fairness} seek to ensure this representation at different ranks. Starting from the target distribution $D$, DetGreedy populates the ranking progressively from the top to the bottom rank with the most relevant items from underrepresented groups. To do so, it maintains a counter $N_g^k$  for the number of items from group $g$ that have already been placed in the top $k$ positions of the ranking; this counter determines two priority sets from which items at rank $k+1$ can be drawn. The high priority set $\mathcal{G}_H$ consists of items from groups that are below the desired quota. The low-priority set $\mathcal{G}_L$ consists of groups that are above their desired quota, but only by a small margin.
\begin{align*}
    \mathcal{G}_H &= \{i \in g: N_g^k < \lfloor D_g \cdot (k+1) \rfloor \} \\
    \mathcal{G}_L &= \{i \in g: \lfloor D_g \cdot (k+1) \rfloor  \leq N_g^k < \lceil D_g \cdot (k+1) \rceil \} \\
\end{align*}
If $\mathcal{G}_H$ is not empty, DetGreedy chooses the most relevant item from the groups in $\mathcal{G}_H$; otherwise, it samples one from $\mathcal{G}_L$.
\begin{align*}
    \tau(k+1) = \begin{cases}
    \text{argmax}_{i \in \mathcal{G}_H, i \notin \{ \tau(1), \dots, \tau(k) \}} f_{\text{soft}}(x_i),& \text{if } \mathcal{G}_H \setminus \{ \tau(1), \dots, \tau(k) \} \neq \emptyset \\
    \text{argmax}_{i \in \mathcal{G}_L, i \notin \{ \tau(1), \dots, \tau(k) \}} f_{\text{soft}}(x_i),              & \text{otherwise}
\end{cases}
\end{align*}

By prioritizing items with higher scores in $\mathcal{G}_L$, DetGreedy may end up violating some minimum representation constraints at some rank, i.e. $\exists g \text{ s.t. } N_g^k < \lfloor D_g \cdot k \rfloor$. This happens when more than one group is in $\mathcal{G}_H$. To mitigate this risk, \citet{geyik2019fairness} propose two variants termed \textbf{DetCons} and \textbf{DetConsSort}. DetCons tries to avoid this occurrence by prioritizing groups that are more likely to enter $\mathcal{G}_H$ at the next iteration, 
while DetConsSort is a non-greedy algorithm that can re-order previous items dynamically to avoid constraint violations at the current rank. DetGreedy is implemented in \emph{LinkedIn Recruiter} to ensure \wasreplace{equity in gender representation}{equitable gender representation in \textbf{candidate search}}; the desired gender distribution $D_g$ is made query-dependent, and set to match the distribution of qualified candidates for the search criteria. 

\textbf{CDF Rescoring} \citep{nandy2022achieving} is a post-processing method targeting $\xeo$ for \wasnew{\textbf{recommender systems}}. $\xeo$ is a fine-granularity measure defined in Equation \eqref{eq:xeo}, studying the properties of the soft classifier $f_{\text{soft}}(x)$ at every possible threshold; it requires that the probability of positive items in a group $\{i \in g: y_i=1\}$  achieving a score below a threshold $t$ should be the same for every group, at every threshold $t$. This is achieved by re-mapping item scores to their groupwise Cumulative Distribution Function (CDF) as:
\begin{align*}
    f'_{\text{soft}}(x_i) = \Pr(f_{\text{soft}}(x) \leq f_{\text{soft}}(x_i)|y=1, s=s_i)
\end{align*}
\noindent In other words, new soft scores are mapped to the $[0,1]$ interval, according to the CDF of old scores for positive points in their group, ensuring $\xeo=0$. \citet{nandy2022achieving} propose several extensions to this approach, to achieve $\xeo$ across all target class values, to account for position bias in the target, and to trade off fairness and accuracy. This method is tested on a \wasnew{\textbf{friendship recommendation}} engine in a live proprietary system (most likely \emph{LinkedIn}), where sensitive groups are defined based on the level of activity on the platform, mitigating potential differences in platform engagement.

\textbf{Spatial Partitioning} \citep{burke2021fair} is a heuristic to select an optimal group of applicants $\sigma$ from screening tests, with the additional difficulty that sensitive attribute values are unknown during testing. First, two estimators for the target ($f_{\text{soft}}(x)$) and protected variable ($h_{\text{soft}}(x)$) are developed on a training set. These estimators are applied to the test set, which is ranked customarily as $\tau = \text{argsort} (f_{\text{soft}}(x))$.  The top candidates are selected from the ranking and put into $\sigma$. Given the systematic groupwise differences encoded in the data, the selected candidates tend to be mostly from the privileged group. Spatial Partitioning mitigates this bias by replacing the candidates in $\sigma$ who have the lowest values of $[f_{\text{soft}}(x)+h_{\text{soft}}(x)]$--or some other linear combination of the two--with the candidates who have the highest values of $[f_{\text{soft}}(x)+ h_{\text{soft}}(x)]$ among the ones that were originally not chosen. This approach aims to rebalance the privileged and disadvantaged group in the final set $\sigma$ while maintaining good accuracy in the selection of high-performance candidates.


\subsection{Discussion}\label{subsec:discussion}

\new{In this section, we present the main trends for fairness enhancement in the algorithmic hiring literature and describe important factors guiding these choices (summarized in Table \ref{tab:miti}). Section \ref{sec:practice} will expand on this analysis presenting additional dimensions that guide fairness mitigation in practice.}


\namedpar{Mitigating biases}.
\new{Some of the proposed approaches are suited to mitigate specific biases described in Section \ref{sec:bcf}.  Balanced sampling, for example, can counter the effect of under-representation in training sets caused by factors such as horizontal and vertical job segregation. Proxy reduction methods (e.g. decorrelation, adversarial inference) target sensitive attribute proxies, reducing the overreliance of models on sensitive information.  Group norming seems applicable to counter systematic biases encoded in input features against protected groups, such as biased employee evaluations. However, it is certainly preferable to understand and improve the measurement system that provides these biased features rather than simply matching the distribution of input features across protected groups without a clear understanding of the underlying socio-technical system.  Finally, output re-ranking can mitigate the joint effect of different bias conducive factors. Its application comes with risks and opportunities described below.
}

\namedpar{Opportunities and risks of post-processing}. Post-processing approaches\new{, such as DetGreedy,} are the easiest to integrate into existing systems, as they can be modularly added to algorithmic pipelines \citep{vasudevan2020lift}. Indeed, most post-processing methods in algorithmic hiring are proposed and deployed at \emph{LinkedIn}, a large company with an established platform powered by interactions between complex data infrastructure and interdependent algorithmic modules \citep{geyik2018talent,aditya2012data}. It is worth noting that post-processing explicitly takes into account sensitive attributes to change algorithmic outcomes for job candidates, which may be critical for disparate treatment and affirmative action (US) as well as direct discrimination and positive action (EU) \citep{bent2019algorithmic,hacker2018teaching}. Additionally, postprocessing requires run-time access to the sensitive attributes of all data subjects\new{, as described in the next paragraph}. These \wasreplace{is likely a key factor}{are likely two key factors} explaining why post-processing is less popular and why hybrid approaches combining different types of fairness interventions, e.g. mitigating via pre- \emph{and} post-processing,  remain under-explored \citep{arafan2022end}.

\namedpar{Sensitive data}. \new{Different mitigation strategies have different requirements for sensitive attributes $s$. Post-processing methods, such as DetGreedy and CDF restoring, typically require knowledge of sensitive attributes during runtime. Conversely, pre-processing and in-processing approaches, such as importance-based scraping and adversarial inference, require access to sensitive attributes only during training, not testing. Furthermore, specific data modalities, such as text and vision, facilitate specialized methods like subspace projection and face decorrelation, which function without sensitive attribute information about data subjects. Each of these methods has distinct data requirements, ranging from highly restrictive to more lenient. Having access to sensitive attributes at runtime enables precise and reliable interventions, whereas strategies that do not rely on this data must be continuously monitored to ensure they are meeting their intended goals.}  

\namedpar{Outcome fairness through \wasreplace{process fairness}{proxy reduction}}? A very clear trend in the literature is the popularity of proxy reduction methods\new{, such as adversarial inference,} that target process fairness ($\sauc$) in pursuit of outcome fairness (e.g. $\di$, $\tprd$). This approach probably gained popularity because it aligns with legislation against \emph{disparate treatment} (US) and \emph{direct discrimination} (EU), which prohibit basing hiring decisions on protected attributes \citep{lambrecht2019algorithmic,adams2023directly} \wasnew{and because it does not require run-time access to sensitive attributes}. 
This approach is also adopted by vendors, such as \emph{Pymetrics} and \emph{Hirevue}, upon detecting violations of the 80\% rule in pre-deployment audits \citep{raghavan2020mitigating,hireview2022explainability}. While this method can mitigate large outcome disparities, it does not produce convincing improvements when the inequity is less significant \citep{hemamou2021judge,hemamou2022delivering,booth2021bias}. Moreover, it is worth noting that, in contrast to post-processing, this approach achieves mitigation on a held-out test set during development, but provides no guarantee at deployment. More general studies are required to understand the effects of proxy removal on outcome fairness and ensure actual benefits for vulnerable candidates.

\namedpar{The problem with videos}. Algorithmic screening can take advantage of data from multiple sources, including gameplay, psychological assessments, and video interviews. The latter provide three types of signals, namely visual, verbal, and paraverbal. Across multiple studies, systems trained on video signals are shown to yield the largest disparities and disadvantages for protected groups \citep{hemamou2021judge,yan2020mitigating,booth2021bias}. Visual signals have been removed from several products \citep{maurer2021hirevue} because they inevitably encode sensitive information such as race and gender while lacking a solid foundation to justify their use in the hiring domain. Even if found to be accurate and fair in a specific evaluation, hiring algorithms based on face analysis are unlikely to predictably generalize and maintain accuracy or fairness under variable conditions, or at least they are less likely to do so than algorithms based on more established data modalities and better-understood correlations with job performance. Furthermore, it is worth noting regulation proposals against inference of emotions, states of mind, or intentions from face images in the workplace \citep{ep2021proposal,ep2023artificial}.
Given this evidence, we invite particular caution against new proposals of hiring systems based on computer vision, advertised as capable of inferring the motivation \citep{kappen2021objective} and personality traits \citep{mujtaba2021multi} of candidates, even if accompanied by bias mitigation and fairness evaluation.  


\section{Data}\label{sec:data}

Datasets used in algorithmic fairness research for the hiring domain are summarized in Table \ref{tab:datasets}. The following sections present these resources divided into textual, visual, and tabular datasets. \wasnew{In line with recent literature, we find no graph dataset on algorithmic hiring \citep{dong2023fairness,chen2023fairness}.}

\subsection{Text-based datasets}


Textual datasets consist of job descriptions and job seekers' resumes or biographies, focusing on professional experience, training, and skills. Resumes and biographies contain strong proxies for sensitive attributes such as race and gender, including names and addresses.

\textbf{Chinese Bios} is a textual dataset generated by \citet{zhang2022are} to study gender bias in candidate rankings produced by BERT-based resume retrieval systems \citep{devlin2018bert}. It consists of short resumes containing a binary gender indicator (he/she) and a two-sentence description of job skills for IT, finance, and administration. 

\textbf{Bias in Bios}, \citep{dearteaga2019bias} 
is composed of textual biographies written in English and extracted from the Common Crawl dataset. It was initially proposed to study fairness in occupation classification. Gender is automatically extracted based on the use of pronouns in the short biographies. Professions are the target variables; they are self-reported in the descriptions. 

\textbf{Engineers \& Scientists} \citep{cowgill2018bias} results from a field study comparing human and algorithmic CV screening in an unspecified company. It is composed of applications, i.e. pairs of resumes and job postings for engineers (software and hardware) and technical scientists. The features were parsed from resumes, including candidate education (institutions, degrees, majors, awards), work experience (job titles and companies), skills, and relevant keywords. The target variable indicates whether the candidates were interviewed and extended an offer.

\textbf{IT Resumes} \citep{parasurama2022gendered2} was used to study how men and women describe themselves on resumes and whether the difference impacts hiring outcomes. The dataset comprises approximately 900,000 resumes (without names, emails, and URLs) in the historical hiring records of eight IT firms based in the US, relevant to just over 6,000 job postings from the IT sector. The resumes were managed through an applicant tracking system, where the applicants self-reported their gender. The target variable encodes whether candidates received a callback after applying.  

\textbf{CVs from Singapore} \citep{deshpande2020mitigating} was introduced to investigate ethnicity bias in automated resume filtering. It contains 135 resumes of candidates of Chinese, Malaysian, and Indian origin (the predominant ethnic groups in Singapore)  who applied for vacancies in Singapore's accounting and finance sectors.
The dataset curators annotated candidate ethnicity based on geographical information on their education and early employment. For example, candidates who completed their education in China were classified as ethnically Chinese. Nine job postings describing open positions in the financial sector are considered. Three annotators annotated each posting-resume pair with a binary variable indicating whether candidates appear qualified for a job based on their CV. 

\textbf{DPG Resumes} \citep{rus2022closing} contains over 10 million vacancies (including salary range, working hours, and job descriptions) and just under 1 million resumes augmented with job categories of interest and gender inferred from first names. Given the imbalance between vacancies and candidates, this dataset is suitable for studying recommender systems that propose jobs to candidates. The data is provided by DPG Recruitment in anonymized form, removing all names (including company names), dates, addresses, telephone numbers, email addresses, and websites. 


\subsection{Visual datasets} 

Several datasets in the hiring space are multimodal, with a strong focus on videos of candidates answering job interview questions in front of a camera \citep{booth2021bias,kappen2021objective,singhania2020grading}, often called Automated Video Interviews (AVI). This data modality is relatively new in the hiring domain. Similarly to CVs, AVIs encode much information on sensitive attributes, including gender, race, age, and disability.

\textbf{ChaLearn First Impression} \citep{escalante2020modeling} contains 10,000 YouTube video clips of people facing a camera and speaking in English.  Amazon Mechanical Turk workers were hired to annotate each clip with the personality traits of the speaker (openness, conscientiousness, extroversion, agreeableness, neuroticism) and a ``variable indicating whether
the subject should be invited to a job interview or not''. The gender and ethnicity of the speakers were annotated by two dataset curators. \todo{This dataset is quite popular, yet very far from something representative}

\textbf{Student Interviews} \citep{booth2021bias} consists of video interviews with 733 upper-level undergraduate students, who were asked to participate in a simulated interview answering six questions. Verbal (e.g., n-gram frequencies), paraverbal (e.g., loudness, jitter, shimmer), and visual (e.g. facial expressions, body motion) features were automatically extracted from videos using tools like OpenSmile \cite{eyben2010opensmile}, FACET \cite{stockli2018facial}, or Motion Tracker \cite{westlund2015motion}. Data were annotated by three research assistants who assessed the candidates' ``hireability'' on a 5-point Likert scale. Gender is self-reported, including a non-binary option.

\textbf{SHL Interviews} \citep{singhania2020grading} is a proprietary AVI dataset with more than 5,000 videos from 810 real job seekers from the US, UK, India, and Europe answering behavioral and domain knowledge questions. Videos were annotated by at least five assessors, who rated the presence of four social skills: engagement, positive emotion, calmness, and confidence, using a scale of 0 to 4. The sensitive attributes available with the dataset are country, age, gender, and race. 

\textbf{Oil Company Interviews} \citep{kappen2021objective} was curated to study the problem of predicting candidate motivation in job selection processes. It comprises AVIs with 154 students from Utrecht University carrying out a mock interview with a fictitious oil company. The participants self-reported their motivation (\textit{``To what degree are you motivated to work for the company''}) on a 10-point Likert scale, which represents the target variable. Software tools like OpenFace and EMFACS \cite{ekman2003unmasking} were used to automatically create facial marker features and extract emotions from videos.

\textbf{FairCVs} \citep{pena2020bias} is a synthetic CV dataset combining short bios from \emph{Bias in bios} \citep{dearteaga2019bias}, face images taken from the \textit{DiveFace} database \cite{morales2020sensitivenets}, and numerical features emulating desirable aspects such as availability or previous experience. Artificial target scores are generated for each CV, as a linear combination of numerical features; biased scores are derived from these with an ethnicity- and gender-dependent additive penalty emulating biases in the data. The dataset has been employed to study the extent to which sensitive attribute proxies can contribute to discriminatory models when the target variables on which they are trained exhibit biases against certain protected groups.

\begin{table}[t]
\tiny
  \caption{Datasets used in research on fairness in algorithmic hiring.}
    \label{tab:datasets}
  \begin{threeparttable}
  \centering
  \begin{tabular}{|m{1.5cm}||m{1.1cm}|m{0.7cm}|m{1.1cm}|m{1.1cm}|m{3cm}|m{1.6cm}|m{0.8cm}|}
  \hline
    Dataset name & Used in & Type & Language & Geography & Target variable & Sensitive variable* & Hiring stage \\ \hline \hline
    Chinese Bios & \cite{zhang2022are} & Textual & Chinese & Synthetic & Mention of job-specific skills in bio & Gender (B, A) & Sourcing \\ \hline
    Bias in Bios & \citep{dearteaga2019bias, hemamou2022delivering} & Textual & English & World & Candidate occupation & Gender (B, A) & Sourcing \\ \hline
    Engineers \& Scientists & \cite{cowgill2018bias} & Textual & English & Unknown & Candidate received an offer & Unknown & Screening \\ \hline
    IT Resumes & \citep{parasurama2022gendered2,parasurama2021degendering} & Textual & English & US & Candidate was called back & Gender (B, S) & Sourcing \\ \hline
    CVs from Singapore & \cite{deshpande2020mitigating} & Textual & English & Singapore & CV match with job description & Ethnicity (A) & Sourcing \\ \hline
    DPG Resumes & \cite{rus2022closing} & Textual & English, Dutch  & Netherlands & Candidate industry of interest  & Gender (B, A) & Sourcing \\ \hline
    ChaLearn First Impression & \cite{hemamou2021judge,yan2020mitigating,kochling2021highly} & Visual & English & Unknown & Speaker hireability and personality annotated by AMT workers   & Gender (B, A), ethnicity (A) & Screening \\ \hline
    Student Interviews & \cite{booth2021bias} & Visual & English & Unknown & Speaker hireability annotated by research assistants & Gender (NB, S) & Screening \\ \hline
    SHL Interviews & \cite{singhania2020grading} & Visual & English & US, UK, India & Speaker communication skills rated by experts & Gender (B), age, race, country of residence  & Screening \\ \hline
    Oil Company Interviews & \cite{kappen2021objective} & Visual & Dutch & Netherlands & Candidate self-reported commitment & Gender, age & Screening \\ \hline
    FairCVs & \cite{pena2020bias,hemamou2022delivering} & Visual & English & Synthetic & Synthetic score & Gender (B, A), race (A) & Sourcing \\ \hline
    Requirements and Candidates & \citep{markert2022} & Tabular & Synthetic & Synthetic & Synthetic score & Synthetic & Sourcing \\ \hline
    Jobs and Candidates & \citep{arafan2022end} & Tabular & Unknown & Netherlands & Candidate recruited or shortlisted & Gender (B, S) & Sourcing \\ \hline
    Pymetrics Bias Group & \citep{wilson2021building} & Tabular & Unknown & Unknown & Similarity to current employees & Gender (S), race (S) & Screening \\ \hline
    IBM HR Analytics & \citep{ghazimatin2022measuring} & Tabular & Unknown & Unknown & Employee resignation & Gender (B), age, marital status & Evaluation \\ \hline
    Resume Search Engines & \citep{chen2018investigating} & Tabular & English & US & Unknown & Gender (B, S) & Sourcing \\ \hline
    Walmart Employees & \cite{burke2021fair} & Tabular & English & US & Employee tenure and performance ratings & Synthetic & Screening \\ \hline
    Web Developers Field Study & \cite{avery2023does} & Tabular & English & US & Unknown & Gender (B,S), age (S), ethnicity (S)  & Screening \\ \hline
    Chinese Job Recommendations & \cite{zhang2022understanding} & Tabular & Chinese & China & Unknown  & Gender (B,A), age (A)  & Sourcing \\ \hline
    Facebook Ads Audiences & \cite{ali2019discrimination} & Tabular & English & US & User clicks & Gender (B, S), race (B, A)  & Sourcing \\ \hline
    \end{tabular}
    \begin{tablenotes}
            \tiny
            \item * where reported, we indicate the provenance of sensitive attributes as annotated (A) or self-reported (S); we also indicate the gender encoding as binary (B) or
non-binary (NB).
        \end{tablenotes}
  \end{threeparttable}
\end{table}

\subsection{Tabular data}
\label{sec:data_tab}

Tabular datasets encode structured data of a diverse nature, describing job-seekers and employees at different stages of the algorithmic hiring pipeline.

\textbf{Requirements and Candidates} \citep{markert2022} is a synthetic dataset with numerical and boolean values encoding both candidate skills and job requirements. Bias against certain candidates is deliberately introduced in the data through additive noise.

\textbf{Jobs and Candidates} \citep{arafan2022end} is a tabular dataset describing candidates and job postings with real-valued, categorical, and binary features. For candidates, their education, experience,  preferences (minimum salary, preferred working hours, maximum travel distance), and self-reported gender are included. Regarding job postings, the dataset contains information about the industry, company size, and geographical location. Joint candidate-job features describing overlaps, such as the distance of candidates' residence from the place of work, are also present. The target variable indicates whether a candidate was recruited or short-listed for a job. 

\textbf{Pymetrics Bias Group} \citep{wilson2021building} is a test set used for pre-deployment audits at Pymetrics, a company offering gameplay-based screening tools to clients. The gameplay of job seekers and current employees of the client company are compared to find candidates who resemble high-performing incumbent employees. The covariates consist of gamified psychological measurements; sensitive attributes, which can be self-reported on a voluntary basis, include gender and race.

\textbf{IBM HR Analytics}\footnote{Available at \url{https://github.com/IBM/employee-attrition-aif360}.} is a synthetic dataset curated by IBM data scientists to study employee resignation, which is the target variable.  Covariates include education, job satisfaction, income, years of service in the company, and commuting distance, along with sensitive attributes such as gender, age, and marital status.

\textbf{Resume Search Engines} \citep{chen2018investigating} includes search results crawled from employment websites \textit{Indeed}, \textit{Monster}, and \textit{CareerBuilder} for 35 job titles in 20 cities of the US, collecting data on 855,000 job candidates. Data were crawled in 2016 to study gender biases.

\textbf{Walmart Employees} was released as part of the Society for Industrial and Organizational Psychology machine learning competition\footnote{\url{https://eval.ai/web/challenges/challenge-page/527/overview}} to study the problem of predicting employee retention and performance from pre-employment tests with questions on work history, personality, and behavioral scenarios. Target variables encode employee tenure and performance ratings. Each instance has a synthetic binary variable that mimics a protected attribute.

\textbf{Web Developers Field Study} \citep{avery2023does} summarizes the results of an experiment on the impact of algorithmic hiring tools on gender diversity. The curators advertised a web developer position for US residents; upon applying, candidates provided information on their education, experience, and demographics, as well as free-form responses to selected questions. The responses were rated using a recruitment tool with a score of up to 100. Each applicant was then rated by a human assessor based on experience and education; assessors were divided into three groups based on whether they had access to algorithmic scores and candidate names.  

\textbf{Chinese Job Recommendations} \citep{zhang2022understanding} consists of job recommendations from four Chinese boards to fictitious profiles that differ only by gender. Profiles are accessed programmatically every two weeks to record job recommendations, which are then compared between different genders.

\todo{Include RecSys Challenge 2016 + 2017 Datasets?}

Several tabular datasets have been curated to measure discrimination in job ad delivery \citep{ali2019discrimination,lambrecht2019algorithmic,imana2021auditing}, using a common methodology. \textbf{Facebook Ads Audiences} \citep{ali2019discrimination} exemplifies this methodology by running a job ad campaign on Facebook and studying differential delivery along gendered and racial lines. They design advertising creatives (headline, text, and image) for different professions, using them in a campaign that optimizes clicks, which are then broken down into different demographics. Despite the fact that the Facebook Marketing API does not allow a breakdown by race, the curators perform this analysis with a careful design of target audiences leveraging phone numbers and racial information available from North Carolina voter records.

\subsection{Discussion}


\namedpar{Low diversity}. English is by far the dominant language; datasets with geographical information primarily represent US citizens (6 datasets) and the Netherlands (3 datasets). We offer two interpretations of these findings. On the one hand, they reflect the importance of both countries in the recruitment industry.\footnote{\url{https://www2.staffingindustry.com/eng/Editorial/Daily-News/World-World-s-largest-staffing-firms-post-224-billion-in-revenue-56012}} On the other hand, they are consistent with previous research reporting that most efforts to improve fairness in artificial intelligence are influenced by the Global North \cite{okolo2022making, roche2021artificial}, particularly the US \cite{Jobin2019}, and focus on English language resources \cite{ramesh-etal-2023-fairness}.

\namedpar{Missing stages}. The vast majority of datasets describe the early stages of the hiring pipeline, i.e. sourcing (11 out of 20) and screening (8 out of 20). We found no datasets (and consequently no studies) for selection, and only one for evaluation (\textit{IBM HR Analytics}). This is expected given the industry's tendency to use algorithms primarily for sourcing and screening \cite{hmoud2019will, li2021algorithmic}. 
However, given the growing push to adopt this technology in later stages \citep{verma2023ai}, there is a tangible risk that algorithms for selection and evaluation will be quietly deployed without a clear understanding of risks and limitations. Indeed, datasets that target employee tenure and performance, such as \textit{Walmart Employees} and  \textit{IBM HR Analytics}, signal an interest from companies in understanding the factors that predict future productivity and loyalty. 

\textbf{Lacking sensitive attributes}. Most importantly, attributes such as disability, religion, and sexual orientation are simply missing, despite the special legal status of these attributes and the evidence of workplace discrimination  \citep{ozeren2014sexual,padela2016religious,ameri2018disability}. Gender is by far the most common sensitive attribute, overwhelmingly encoded as binary. Ethnicity and race are considered in 6 out of 20 datasets, making this the second most common attribute. They are frequently annotated, rather than self-reported, in textual and visual datasets, \wasreplace{confirming that}{due to} these data modalities \wasreplace{carry}{carrying} strong proxies for sensitive attributes. 

\textbf{Target multiplicity}. HR management and recruitment tasks allow multiple formulations. Many target variables can seem reasonable at face value; therefore, initial data curation and design choices have a prominent role. As candidates move through the hiring pipeline, their digital record goes through a data journey where they are marked as aware of a position, applying (or headhunted), proposed to a client, screened-in, interviewed, hired, retained, promoted, and so on. Companies interested in algorithmic hiring solutions pick one or more of these variables, balancing different priorities such as efficiency and quality of hire, with unpredictable effects on algorithmic fairness, a phenomenon called \emph{multi-target multiplicity} \citep{watsondaniels2023multi}. Indeed, the target variables in Table \ref{tab:datasets} are very diverse. On the one hand, this reflects the length of hiring pipelines and the diversity of data journeys. On the other hand, 
it points to a lack of established best practices around target variables. Focusing on screening datasets, we find different constructs (e.g. communication skills vs. commitment)  annotated by people with different competencies (AMT workers vs. experts) from disparate data sources (YouTube videos vs. mock interviews). Since the validity of these estimates and their connection with job performance has been called into question \citep{barrick1991big,rhea2022external}, we call for caution in handling these variables and granting them the status of \emph{ground truth}. This can be especially problematic when using conditional fairness measures whose very definition hinges on this so-called ground truth.

\section{Measurement and Mitigation in Practice}
\label{sec:practice}
\wasnew{In this section, we present practical considerations on bias mitigation and measurement that emerge from our review and from direct involvement in the industry. We focus on key technical factors guiding measurement and mitigation choices.}

\subsection{Data modalities}
\wasnew{First, different data modalities enable different bias mitigation strategies.}

\wasnew{\textbf{Textual data} is a common data modality encountered in the sourcing stage, typically in the form of textual CV or job description data, or in the screening stage, e.g. transcripts of video interviews.
Common bias mitigation methods that cater exclusively to textual data include rule-based scraping or dictionary-based methods. 
These methods, also offered by vendors such as \emph{Textmetrics} and \emph{Textio}, can be used for mitigating specific types of biases, e.g., age bias in job descriptions~\cite{fokkens2018leeftijdsdiscriminatie}, or gender bias in biographies~\cite{dearteaga2019bias}. 
These mitigation strategies tend to be technically straightforward to implement, as they can be developed as standalone components and need not be integrated in complex algorithmic hiring pipelines. 
In addition, collecting the required data (e.g. constructing task-specific dictionaries) leans heavily on domain and context-specific knowledge, which is typically available to practitioners in the HR domain.} 
\wasnew{A drawback of these dictionary-based methods is a different side of the same coin: due to reliance on domain, context, and task-specific data (dictionaries) they do not transfer well over different problems (e.g., scraping gendered words from resumes vs. substituting words associated with age discrimination in job descriptions require wholly different sets of terms), languages, or geographies (where cultural or regulatory differences may impose different requirements).}

\wasnew{Additionally, when it comes to textual bias mitigation, the rapid uptake of LLMs has pulled into focus biases that arise from LLM-based natural language generation. 
For example, \citet{salinas2023unequal} show how LLMs exhibit gender and nationality bias when generating job recommendations for job seekers of different genders and nationalities. 
They find that the types of jobs recommended follow common gender and nationality-based stereotypes. 
In a similar experiment, GPT-4 is found to more frequently use female pronouns in reference letters generated for female-dominated occupations (such as nannies), and male pronouns for male-dominated occupations (such as plumbers)~\cite{bubeck2023sparks}. 
A simple yet effective mitigation strategy here is prompt engineering: by appending the phrase \emph{``in an inclusive way''} to the prompt, previously gendered pronouns are replaced with third person pronouns (``they/their'').}
\wasnew{While novel bias mitigation strategies for LLMs are actively studied~\cite{li2023survey}, the previous examples show that being mindful of which information to include in a prompt (either explicit or implicit), and understanding how to effectively construct prompts are important first mitigation steps for leveraging LLMs in the context of algorithmic hiring.}


\wasnew{Next, \textbf{visual data}, either through images or video, enable and require a different set of bias mitigation methods and strategies. 
As shown in Section~\ref{sec:miti}, video-based systems tend to yield more disadvantageous effects for protected groups, also due to the strong encoding of sensitive information. 
In this light, \emph{EASYRECRUE} present an adversarial method that removes sensitive information from latent representations of neural networks that were trained for predicting `hireability' given (features that represent) facial expressions of candidates in job interviews~\cite{hemamou2021judge}. This type of adversarial bias mitigation can, however, be applied over a wider set of data modalities.
In the end, due to video-based algorithms unreliability in the face of varying conditions, and incoming regulation proposals that prohibit inference of emotions or intentions from facial data in workplace contexts (discussed in more detail in Section~\ref{subsec:discussion}), it is advisable for practitioners to approach video-based hiring systems and tools with caution, irrespective of which mitigation strategy to choose.}

\wasnew{Finally, \textbf{tabular data} is a common data modality in algorithmic hiring and machine learning more broadly. 
Mitigating bias in tabular data can be done through pre-processing approaches that directly intervene on the features, e.g. by removing features that are highly correlated with sensitive attributes via importance-based scraping, or increasing the representation of vulnerable populations in training sets with balanced sampling.
These types of methods are widely available in open-source bias mitigation software packages such as the \emph{Fairlearn} Python package~\cite{weerts2023fairlearn} or the \emph{AI Fairness 360} toolkit (available in Python and R)~\cite{aif360-oct-2018}. 
In addition, we see practitioners experiment with alternative pre-processing methods in algorithmic hiring, such as synthetic tabular data generation~\cite{SDV} for measuring~\cite{vanels2022improving} and mitigating bias~\cite{arafan2022end}.}

\subsection{Tasks}
\wasnew{Next to different types of data modalities, different downstream tasks can enable and call for different measurement and bias mitigation strategies. 
The two most common tasks in algorithmic hiring are classification and ranking.}

\wasnew{For classification tasks, getting started with bias mitigation is relatively straightforward, as there exist several open source or otherwise freely available software packages and libraries, as mentioned above, that provide different implementations of bias measurement methods and mitigation algorithms, specifically designed for classification tasks. 
}

\wasnew{Ranking is a common task in the sourcing and screening stages of hiring, performed by search engines or recommender systems. 
However, the prevalence of classification-based bias metrics in algorithmic hiring (discussed in more detail in Section~\ref{sec:measures}) is reflected in the aforementioned open source packages, which means that measuring and mitigating bias in ranking systems tends to follow the practice of re-purposing classification methods through, e.g., thresholding rankings (i.e., cutting off at $k$). We recommend using fair ranking measures \replace{whenever possible for the evaluation of ranking systems, although their adoption can be complicated by the need to model browsing behavior}{for an evaluation of ranking systems that is more cognizant of user browsing behavior, although, of course, the latter needs to be suitably modeled}. While several mitigation methods apply to both classification and ranking tasks (e.g. adversarial inference), practitioners should be aware of ranking-specific methods such as DetGreedy.
}


\subsection{Scalability and efficiency}
\wasnew{Technical and infrastructural decisions further affect which bias mitigation methods to consider. 
For example, \citet{geyik2019fairness} decouple their post-processing method from specific model choices and properties of input data, which means their method can naturally scale across the different (ranking) systems at \emph{LinkedIn}. 
This also means that such a method can be developed and deployed as a standalone component or micro-service, which decreases development time, effort, and alignment, when compared to pre-processing or in-processing methods that may need to be designed and integrated in complex algorithmic hiring pipelines. In general, post-processing approaches offer an advantage in terms of system integration, but they should be applied with care due to anti-discrimination law (Section \ref{sec:legal}).
}

\wasnew{In terms of algorithmic efficiency, \citet{geyik2019fairness}'s re-ranking method for bias mitigation can be considered computationally cheap as only a subset of a model's output needs to be processed (i.e., top-$k$ items). 
At the same time, while low, the additional computational costs will be incurred for each ranker output.}
\wasnew{Bias mitigation through pre-processing, such as counterfactual data augmentation, and in-processing, such as feature weighting, can be comparatively more resource intensive as they involve (re-)training models. 
Adversarial inference methods, which have been applied to a variety of data modalities and tasks, even require training multiple neural networks simultaneously, which means they can incur substantially higher costs compared to their non-mitigated counterparts. 
However, while these additional costs may be incurred at training time, the resulting models do not incur additional costs at inference time.} 
\wasnew{In the end, algorithmic efficiency and computational costs will vary across bias mitigation methods in nature and magnitude, with simple rule-based scraping methods that involve dictionary look-ups on the cheap end of the spectrum, and bias mitigation methods that involve re-training LLMs at the resource-intensive end~\cite{li2023survey}. 
Aspects such as task, model complexity, architecture, training parameters, dataset size, and composition influence the scalability and efficiency of bias mitigation strategies.}

\subsection{Sensitive attribute data availability and usage}
\wasnew{Different bias mitigation methods have different requirements around the availability of sensitive data, which can further steer mitigation method selection.}

\wasnew{The availability of sensitive data is affected by several factors in practice. 
First, as we identify in Section~\ref{sec:legal}, access to sensitive attributes may be at odds with privacy regulations such as the GDPR, which is an important real-world constraint. 
Next, availability of and access to sensitive attributes may require job seekers' explicit consent, which can be difficult to acquire at the scale needed for some bias mitigation methods. 
Finally, some sensitive attributes, such as age and gender, may have to be recorded in the hiring process for identification purposes, and can hence be assumed to be available for all job seekers. Many other sensitive attributes (e.g., religious beliefs) will not be easily available, and will have to be explicitly requested for bias measurement and mitigation purposes.}

This lack of access to sensitive attributes means some bias mitigation methods may be less suited than others\new{, as noted in Section \ref{subsec:discussion}}; in particular, post-processing methods that require sensitive attributes for all candidates at inference time can be unrealistic. 
\wasnew{Here, pre-processing bias mitigation methods may prove more useful, as they (only) require access to sensitive attributes at training time, and can operate on attributes even when available for only a subset of job seekers (e.g., those who provided consent). 
Furthermore, this allows these methods to be deployed and run in isolation from production systems in controlled batch scenarios, which can be another important practical benefit. 
Examples of these bias mitigation methods used by practitioners include adversarial inference~\cite{rus2022closing}, and re-balancing training data with synthetic data generation~\cite{arafan2022end}.}

\wasnew{Finally, a third family of bias mitigation methods that can operate without requiring access to sensitive attributes at all are the rule-based approaches described above, which are commonly used for mitigating bias in textual data such as job descriptions or resumes. 
These methods rely solely on domain and task-specific gazetteers or dictionaries, and hence can be used when access to sensitive attributes of individuals is not available or desirable.}

\subsection{Fairness definitions and intervention targets}
\wasnew{Once data modality, task, infrastructural and technical choices, and access to sensitive data are set, one important practical challenge is that of defining ``fairness'' and formulating the intervention target of a bias mitigation method--i.e. deciding ``when to intervene'' and ``what to optimize for''.}

\wasnew{In the case of \emph{LinkedIn}'s Talent Search~\cite{geyik2019fairness}, the DetGreedy algorithm was implemented to have the ranker's top-$k$ output reflect the gender distribution of job seekers that meet the requirements of a recruiter-issued query, i.e. the desired distribution of the ranking is set to be equivalent to that of the underlying population of job seekers. 
Here, alignment with the company's goals and values, interpretability (or ability to explain), and collaboration across different stakeholders were mentioned as key factors in guiding the eventual target distribution. 
Defining fairness and formulating an intervention target is not a one-off task, as algorithmic hiring components will be deployed in different, product-specific contexts and stages of the hiring pipeline. \citet{candela2023disentangling} present the (evolving) framework used at \emph{LinkedIn} that guides definitions and operationalizations of AI fairness across their different (types of) products.
}

\wasnew{In general, the challenge in defining fairness and formulating intervention targets in algorithmic hiring is exacerbated by the multi-stakeholder nature of the hiring domain, where development teams may need to consult legal and compliance teams, HR professionals and recruiters, product managers, and executives, all of whom may need to be informed or provide input. Indeed, these choices should be guided by ethical, social, and legal dimensions. 
This challenge may explain in part the popularity of disparate impact as a fairness metric, due to its ease of interpretation across a broad and diverse stakeholder group, or the popularity of complying with the EEOC's 80\% rule~\cite{eeoc2015uniform} as a fairness target (adopted e.g. by \emph{Pymetrics}~\cite{wilson2021building}), as it appears, at least superficially, a legally grounded target.}

\subsection{Fairness vs. utility trade-offs}
\wasnew{
In the sourcing or screening stages, depending on the intervention target, \emph{outcome fairness} may come at the cost of utility; this can happen, for example, when optimizing for gender parity in heavily male or female-dominated industries. 
However, perhaps surprisingly, different experiments by practitioners show that outcome fairness does not need to come at the cost of utility. 
First, in their online A/B-testing experiments, \citet{geyik2019fairness} find their DetGreedy method improves their fairness metric, with no significant impact on utility. 
In addition, \citet{arafan2022end} find that their pre-processing mitigation method of rebalancing training data may even improve utility over non-bias-mitigated methods, in an offline experiment using real data from an international HR company. 
Moreover, with respect to \emph{impact fairness}, \citet{peng2019what} show that ``overcompensation'' (i.e., artificially over-representing a gender in the output of a ranking system) as an intervention strategy for a hiring algorithm can in some cases mitigate human bias further down the hiring funnel. 
%
}
\wasnew{Overall, this section described several practical considerations constraining bias mitigation interventions; desired utility levels are just one dimension, and often not the most restrictive.   
}

\section{Legal Landscape}
\label{sec:legal}



In this section, we describe the main regulations and non-discrimination provisions concerning algorithmic hiring in the EU and the US. We list the main legal sources in both regions, emphasizing the former since it is less cited in the computer science literature on algorithmic hiring. We close this section with remarks on open challenges and practical concerns at the intersection of technology and policy.

\subsection{European law} 

\subsubsection*{Non-discrimination law}
We focus on rules that apply in the whole European Union (27 Member States). In the absence of specific legal rules on non-discrimination in algorithmic hiring, general non-discrimination law applies. 
The right to non-discrimination is protected as a human right in Europe by the European Convention on Human Rights (1950) and the Charter of Fundamental Rights of the European Union (2020). 
The EU also adopted a number of legal acts called \emph{directives}, prohibiting several types of discrimination in different contexts, which Member States implement (give effect to) by adopting national law \citep{krommendijk2023eu}. The four most relevant non-discrimination directives are the following: the Racial Equality Directive (2000),  the Employment Equality Directive (2000),  the Gender Goods and Services Directive (2004),  and the Recast Gender Equality Directive (2006) \citep{council2000council,council2000council2,council2004council,european2006directive}. 
Together, the directives offer protection against discrimination in hiring on the basis of six grounds, also called protected characteristics, or protected attributes: age; disability; gender; religion or belief; racial or ethnic origin; and sexual orientation.
EU law distinguishes two categories of prohibited discrimination: \emph{direct} and \emph{indirect}. 

\emph{Direct discrimination} means that the person is treated less favorably than another on the basis of a protected characteristic, such as ethnicity \citep{council2000council}.  For example, if a company says it will not recruit people of a certain ethnicity, that is an example of direct discrimination \citep{court2008centrum}.  Direct discrimination is always prohibited, with a few narrowly defined and specific examples. For instance, a women’s clothing brand is allowed to hire only female models for its advertising pictures.

The second category of prohibited discrimination is \emph{indirect discrimination}, occurring when a practice is neutral at first glance but ends up discriminating against people of a certain ethnicity (or another protected characteristic) \citep{council2000council}.  For indirect discrimination, the law focuses on the effects of a practice; the intention of the alleged discriminator is not relevant. Hence, even if an organization can prove that it did not know that its algorithmic system discriminated unfairly, that will not help the organization.

There are three elements of indirect discrimination that can be summarized as follows \citep{borgesius2020price,council2000council}.  (1) The practice must be neutral. For example, rejecting job applications coming from a certain postal code would count as neutral. Rejecting applications from people with a certain ethnicity would not be neutral; it would be direct discrimination. (2) This neutral practice puts people with a certain ethnicity (or other sensitive attributes) at a ``particular disadvantage compared with other persons’'.  The word \emph{disadvantage} must be interpreted in a wide way.  (3) There is no objective justification for such practice. The apparently neutral practice is not prohibited if the ``practice is objectively justified by a legitimate aim and the means of achieving that aim are appropriate and necessary’'.

As a hypothetical example, suppose that people with an immigrant background make more spelling errors in job application letters. A cleaning company never hires job applicants for cleaning jobs if the application contains spelling errors. This practice seems neutral at first glance, but results in the rejection of most applicants with an immigrant background. The cleaning company cannot justify this no-spelling-errors rule because people can be good cleaners, even if they make some spelling errors. Therefore, the cleaning company engages in illegal indirect discrimination. However, the situation would be different for a law firm. The main job of many lawyers is writing precise, and often official, documents. The law firm is allowed to reject applications with spelling errors, even if it results in most people with an immigrant background not being hired.

The organization is also responsible if it uses an algorithmic system provided by, for instance, a company. If the algorithmic system turns out to discriminate illegally, the organization using the system is responsible and the victim can sue it for damages, for instance. (Later, the organization could try to sue the AI developer, but the organization remains responsible towards the victim.) In sum, general non-discrimination law applies to new forms of algorithmic discrimination, also if that discrimination happens indirectly or accidentally.  

\subsubsection*{Other relevant law}
We briefly highlight some other relevant laws in the EU. The GDPR (the General Data Protection Regulation) is, roughly summarized, a European-wide statute that aims to protect fairness and human rights when personal data are used. The GDPR is long and detailed. Among other things, GDPR bans the use of \emph{special categories} of personal data (sometimes called sensitive data). These are data on, for example, someone's ethnicity, religion, trade union membership, health, or sexual orientation (article 9 GDPR). There are some exceptions to the ban. For instance, hospitals are allowed to use health data. Another exception is the individual’s \emph{explicit consent}. Generally speaking, consent is not freely given, and thus not valid, if an employer asks a job applicant or employee for their consent, because of the unequal power relation. This GDPR rule makes it difficult to use sensitive data to audit or train algorithmic systems \citep{van2023using}.

There is a proposal for an AI Act in the EU, with many requirements for ``high-risk'’ systems, including algorithmic systems for HR \citep{ep2021proposal}. Developers of high-risk AI systems must, for instance, ensure that the training data are appropriate and do not lead to unlawful discrimination. \wasreplace{The text is still being negotiated, so the final text is unknown.}{At the time of writing, the EU did not officially adopt the AI Act yet, and did not publish the final text yet.}  The proposals in the AI Act also contain a new exception to the GDPR, to enable the use of sensitive data for debiasing algorithmic systems. 

\subsection{US law}
\label{sec:us_law}

\subsubsection*{Federal law}
US non-discrimination law is similar to EU law in many respects, but also decidedly different in others. Like EU law, its sources are spread over different statutes, and case law plays a crucial role. For example, Title VII of the Civil Rights Act of 1964 constitutes a federal law prohibiting employment discrimination based on race, color, religion, sex, and national origin. Significantly, the Equal Employment Opportunity Commission (EEOC) issues nonbinding, but practically important guidelines to interpret Title VII. Other important sources of federal law are the Age Discrimination in Employment Act of 1967, the Americans with Disabilities Act of 1990, and the Genetic Information Nondiscrimination Act of 2008.

For all these acts, US law distinguishes between two fundamental types of discrimination: disparate treatment and disparate impact. They resemble, but do not perfectly mirror, the difference between direct and indirect discrimination in the EU. Importantly, disparate treatment requires not only an adverse action based on a protected attribute, but also the proof of intent on the part of the discriminating individual – unlike the EU variety of direct discrimination \citep{van2023using}.  To actually win in court, an injured person in the spelling mistake example would have to demonstrate that the cleaning agency introduced the spelling requirement with the purpose of treating unfavorably members of their protected group. In practice, this will often be difficult \citep{us1989price}. 

Disparate impact, in turn, closely resembles indirect discrimination. Intent is not required \citep{us1971griggs}.  Rather, the disparate impact doctrine prohibits actions that are seemingly neutral, but significantly disadvantage members of a protected group. In its guidelines, the EEOC suggested that such a disadvantage usually occurs if the chance of a member of the protected group being positively evaluated is 80\% or less than that of a member of the privileged group (so-called 80\% rule or 4/5 rule) \citep{eeoc2015uniform}.  Finally, disparate impact can be justified if there is a legitimate reason for the practice \citep{us1973mcdonnel}, for example business necessity \citep{us1971griggs,barocas2016big}.  Therefore, a law firm could arguably use a model evaluating orthography and grammar to rank candidates even if this disproportionately disadvantages members of one specific protected group. 

Overall, unless intent can be shown, many cases of algorithmic discrimination will be argued under the disparate impact prong. Therefore, many contributions to the literature on technical algorithmic fairness have provided tools to ensure that this rule is not violated at the statistical level \citep{feldman2015certifying,zafar2017fairness,zehlike2020matching}.  However, courts will generally, both in the EU and in the US, look at factors beyond mere scores and numbers to determine whether a legally relevant disadvantage exists \citep{wachter2021fairness,hacker2018teaching}. 

\subsubsection*{Other relevant law}
Apart from the acts mentioned, federal legislation specifically addressing discrimination in AI systems is unlikely to emerge anytime soon. The Algorithmic Accountability Act is stalled in a gridlocked Congress. Hence, several states and municipalities have taken the initiative and enacted AI hiring laws themselves. For example, the city of New York passed a law on automated employment decision tools \citep{nyc2021automated}.  From July 5, 2023, NYC Local Law 144 applies to employers using AI to substantially assist or replace discretionary decision-making in hiring. They are required to conduct and publish impartial bias evaluations by an independent auditor. Furthermore, the state of Illinois enacted the AI Video Interview Act \citep{illinois2020artificial}.  In force since the year 2020, the law requires employers to put candidates on notice, and obtain their consent, before subjecting their video interview to AI analysis. Candidates may request deletion, and employers that rely solely on AI need to collect and report candidates’ race and ethnicity.

Practitioners have to comply with these local rules if their model is applied in these jurisdictions. Finally, further constraints may arise from affirmative action law, particularly if the fairness intervention goes beyond what is necessary to remedy otherwise unjustified discrimination; this is a complex topic in both the US \citep{us2009ricci,bent2019algorithmic,kim2022race} and the EU legal framework \citep{hacker2018teaching,hoch2021discrimination}. 

\subsection{Operational challenges}

This outline of the EU and US legal landscapes provides some normative reference points for practitioners and offers an opportunity to discuss some practical challenges in assessing the legal compliance of algorithms. 

In general terms, an algorithmic system does not cause indirect discrimination or disparate impact if it pursues \emph{legitimate aims} through \emph{appropriate means}. In practice, several questions arise about both elements at the base of this legal principle. Is inferring a candidate's motivation, i.e. an internal combination of their emotions, states of mind, and intentions, a legitimate aim? Moreover, should estimates by algorithms trained on biased ``ground truth'' variables be considered appropriate means?  In the absence of precise guidelines, these questions have to be assessed contextually to each algorithmic system, and algorithmic implementations should be compared, to the best extent possible, to (hypothetical) outcomes under non-algorithmic alternatives \citep{hacker2018teaching}.

Direct discrimination, in turn, is generally much more difficult to legally justify \citep{adams2023directly}. A natural question arises about the compliance of bias mitigation approaches embedded in hiring algorithms. More specifically, are post-processing algorithms legitimate only if they enforce inter-group parity for equally qualified candidates, e.g. targeting $\tpr$ parity rather than $\di$? If so, what quantitative criteria should be applied to assess candidate qualifications? And to what extent is human intervention and a comprehensive assessment of any re-ranking required? Here, the answers depend on affirmative action (US) and positive action (EU) law, which are currently in flux after the US Supreme Court's decision against Harvard's and UNC's affirmative action programs. At a minimum, human oversight of post-processing operations, and an evaluation of its effects both on groups and on individuals particularly worthy of protection (single parents; chronically ill persons), seems advisable.


\section{The Real World is Messy}
\label{sec:messy}

Our previous discussion has focused on how discrimination can be exacerbated by algorithms. Nevertheless, we have striven to point out the ways such discrimination can be mitigated and even how algorithms can be designed to actively discourage discrimination. It should now be clear that the design of technology plays a key role, either in creating discrimination or by reducing it. To overly value the role of technology design in discrimination or anti-discrimination would also be a mistake, however. This is because non-technological factors have been shown, on occasion, to strongly influence decisions and amplify technological bias, as we will explore in this section.

\subsection{Algorithmic Uptake}

During the COVID-19 pandemic, delivery companies such as Amazon, Deliveroo, and DoorDash rolled out algorithmic recruitment systems to avoid the danger of viral contagion in their HR teams, as well as experiment with a new technology that had the potential to save millions of dollars in HR bills \citep{zhang2022examining}. One reason they were able to do this with ease was due to the regulatory environment, which was laxer than usual due to the ongoing emergency conditions. We have explored some of the ethical problems regarding discrimination in previous sections of this article, as well as some of the potential technological solutions to designing algorithms with anti-discrimination in mind. Each of these solutions is a technological response, however, so we need to remain aware of the non-technological or reduced-technological options. 

Although hiring technology can be more ethical than humans and reduce bias in decision-making \citep{kahneman2021noise}, it often reinforces bias and results in distinctly non-ethical outcomes. This means we should at least consider the balance of ethical harms if we reintroduced a non-technological solution. The recent pandemic necessitated social distancing, but once this danger had passed, non-technological hiring procedures could be reintroduced. To understand why this has not happened, we must consider the institutional incentives of tech companies. These companies have invested heavily in their digital hiring technology as suppliers or customers, so are highly motivated to retain it, even once there is no longer a strong need for it.

\subsection{Algorithmic Fairness}

Recalling the example from Section \ref{sec:bcf_intersec}, suppose that appropriate technical solutions have been deployed in hiring algorithms to counter the bias conducive factors against intersectional minorities. Should she seek a new job, JS would have a sizeable probability of being sourced, screened-in, and recruited -- equal to natives (male and non-male alike) with similar skills and experience in the hospitality industry. However, in the presence of an exogenous shock, such as the COVID-19 pandemic, workplaces are heavily affected, with large consequences on hiring and recruitment. The hospitality industry receives a major blow. Food services stop hiring waitstaff and dismiss most of the employees. Schools and childcare services become unavailable. All of a sudden, JS finds herself unemployed, urgently in need of a new job, which is now more difficult to obtain since demand for her skills has decreased and competition has surged. Sourcing algorithms place JS at the bottom positions of rankings, granting her low visibility. Childcare duties demand much time from her due to her gender \citep{sevilla2020baby}. She has fewer connections than native job seekers for recommendation and referral \citep{zhang2021unequal}. Overall, her probability of successfully reaching the end of a hiring pipeline has dropped substantially and more sharply than for privileged groups. In addition, her migration background makes JS less likely to gain support from social safety nets \citep{mar2022racial,che2020unequal}, increasing the urgency of her need and the unfairness of the new status quo.

This section highlights two limitations of algorithmic fairness in hiring. On the one hand, by restricting their scope to a single system at a precise point in time, fairness measurements run the risk of missing the bigger picture, i.e. the broader socio-technical system in which hiring algorithms are embedded, which can change quickly and profoundly. A model deemed accurate and fair in the old context may perform poorly in the new one.  On the other hand, these changes may be difficult to detect and quantify. Fairness evaluations by practitioners on pre-deployment test sets, such as \emph{Pymetrics Bias Group} (Section \ref{sec:data_tab}), may quickly become obsolete. Strong shocks in the hiring domain are an issue for data representativeness more broadly. Fresh data becomes necessary. This is further complicated by changing incentive structures, by the complexity of handling sensitive data, and by the frequent delay between decisions and feedback in hiring.




\section{Opportunities and Limitations}
\label{sec:pros_cons}


Algorithmic hiring benefits from, and contributes to, fairness and anti-discrimination work, as the previous sections have shown. In this section, we summarize the emerging opportunities and limitations, from which we derive a set of recommendations for researchers and practitioners, summarized in Table \ref{tab:summary}.

\begin{table}[t]
\small
\centering
  \caption{Opportunities, limitations, and recommendations for algorithmic hiring and fairness research.}
  \label{tab:summary}
  \begin{center}
    \begin{tabular}{|p {1cm}|p {3cm}|p {4.1cm}|p {4.5cm}|}
      \hline
      & \textbf{Opportunities} & \textbf{Limitations} & \textbf{Recommendations}  \\
      \hline \hline
       Bias \& validity & consider large candidate pools, reduce human biases, and attract minority candidates & risk of encoding individual biases along with inevitable societal biases; invalid target variables & focus on vulnerable populations beyond acceptance rates; study individual fairness and exploratory policies; \wasnew{carefully} scrutinize new \wasreplace{tools}{technologies} \\ \hline
       Broader context & trigger positive feedback loops; consider tech-recruiter collaboration & narrow focus on local outcomes can overlook fairness in entire hiring process; risk of repurposing for termination & center on job seeker impacts; identify leaks in pipelines; design for recruiter-tech interaction; beware of performance and tenure prediction \\ \hline
       Data & support evaluations of diversity and inclusion & reduced geographical, linguistic, and sensitive attribute coverage & design data collection for diversity; develop IP-friendly audits \\ \hline
       Law & apply binding regulation to positively influence industry & legal restrictions on fairness approaches: concerns about discrimination and data protection & multidisciplinary research balancing fairness, privacy, and anti-discrimination; monitor EU AI Act \\ \hline
    \end{tabular}
  \end{center}
\end{table}



\subsection{Bias and Validity}

\namedpar{Opportunities}. Algorithms for hiring can consider large pools of candidates, avoiding the preliminary exclusion of unusual profiles, as often done by human recruiters under time constraints. Under-representation and sampling biases can be mitigated as a result. Algorithmic hiring also has the potential to mitigate biases in imperfect human judgments. The simple fact of using algorithmic decision-making can reduce avoidance by vulnerable candidates \citep{avery2023does}. Fair and trustworthy algorithms can lead to a positive form of automation bias and attract minority candidates.

\namedpar{Limitations}. Data-driven algorithms tend to encode individual and societal biases. Some algorithms are explicitly trained and evaluated to “predict the competency scores candidates would have been given by trained human reviewers” \citep{orcaa2020description}, inheriting individual biases from recruiters. Previous experience is a preeminent feature for assessing candidates. In conjunction with current job segregation, this means that the most important features are inevitably skewed against historically disadvantaged groups. In addition to these biases, the epistemic validity of prediction targets such as candidates' \emph{employability} and \emph{motivation} is questionable. Job performance is famously difficult to define and measure, let alone predict \citep{robotham1996competences,wang2022against}. Algorithms cloaked in objectivity can promote bias while targeting and legitimizing ill-defined quantities.

\namedpar{Recommendations}. Attention should be devoted not only to acceptance rates for vulnerable populations ($\di$) but also to their representation among applicants, as well as their progress \wasnew{downstream of the algorithm and} post-hiring. \wasreplace{For applications, job descriptions and organizational communication can play an important role}{Job descriptions and organizational communication can play an important role in attracting or repelling specific groups}; automation attempts \citep{qin2023towards} should carefully include fairness evaluations. To mitigate biases against unusual candidates, individual fairness and exploratory policies should be studied. Individual fairness measures can surface problematic situations for individuals that may go unnoticed when studying group fairness. Exploratory policies based on partially stochastic mechanisms can provide new information in repeated decision-making scenarios. However, the social acceptability and procedural justice of such policies in the hiring domain remain to be studied. Finally, new technologies, including AVIs and personality prediction, deserve additional scrutiny, especially through the lens of validity theory \citep{rhea2022external,rhea2022resume}. 

\subsection{Broader Context}

\namedpar{Opportunities}. It is worth highlighting that improving fairness does not require completely removing bias. Algorithmic hiring can reduce certain disparities and trigger deflating feedback loops across bias conducive factors. These algorithms do not (and should not) operate autonomously. Effective and equitable hiring can result from a fruitful interaction between technology and recruiters, leveraging complementary strengths. For example, HR professionals are better suited to assess special cases and operate under changing conditions \citep{li2021algorithmic}. 


\namedpar{Limitations}. Most fairness measures are focused on narrow algorithmic outcomes, neglecting the wider socio-technical context around these algorithms. Some of these measures are completely symmetric and consider advantages for vulnerable and privileged groups as equally problematic (which is, however, generally required by the law). Zooming out from single-algorithm evaluations, it is worth noting that outcome fairness at every stage does not guarantee fair outcomes for job seekers throughout the hiring pipeline. Furthermore, fairness for employers is currently missing; two-sided platforms would be well-positioned to study the performance of their algorithms across employers, devoting special attention to small businesses. Finally, it is worth noting that the discovery of patterns that predict job performance for hiring can open the way to models for termination decisions.

\namedpar{Recommendations}. We call for contextualized and integrated evaluations of decision-making processes that go beyond the predictions of a single algorithm. 
This will help to address complex problems, such as identifying leaks in the hiring pipeline that are most critical for vulnerable groups and modeling impacts on job seekers, such as their efforts and benefits. To exemplify, rejected candidates may still benefit to some extent from a specific type of explanation. Moreover, it will be important to better understand the utility derived from these algorithms by different employers, considering recruitment workflows and developing new fairness measures. The prevalent \emph{human vs. algorithm} evaluation framework is of limited utility; to overcome it,  more research on recruiter-machine interaction is required \citep{suhr2021does}, including candidate screening models \citep{alvarez2023initial} leading to more granular measures. Finally, we invite special caution in the development of predictive models for job performance and tenure, due to a risk of exploitation for termination decisions \citep{zielinski2023should}; such an application of algorithms raises even stronger ethical and social concerns, which are only beginning to be discussed \citep{russon2020uber}.





\subsection{Data}

\namedpar{Opportunities}. Algorithmic fairness research is contributing additional analysis into hiring practices from a perspective of diversity and non-discrimination. More data entails more scrutiny and reflection, which can inform organizational frameworks, such as \emph{Diversity, equity, and inclusion}, and scholarly fields, such as applied psychology and economics. 

\namedpar{Limitations}. Research on fairness in algorithmic hiring is based on data with reduced geographical and linguistic coverage. In addition, important sensitive attributes are missing from the data, making it difficult or impossible to evaluate algorithms for specific vulnerable groups. Data and research are constrained by a dual tension with the privacy of data subjects, on one side, and the intellectual property (IP) of companies, on the other side. 




\namedpar{Recommendations}. Practitioners and researchers should seek more diverse data, with greater geographical and linguistic diversity, and better coverage of sensitive attributes that are relevant in hiring but are lacking, such as disability status and sexual orientation. Dedicated initiatives should be undertaken, including optional surveys for job applicants and broader data donation campaigns \citep{bietz2019data}. Innovative auditing protocols should be studied for employers and providers of algorithmic hiring solutions, including IP-friendly data disclosure procedures.

\subsection{Law}


\namedpar{Opportunities}. In practice, binding regulation shapes algorithmic development more than ethical guidelines or self-regulation. Although clear guidance on algorithmic hiring is currently missing, precise requirements set out in future regulation and case law, informed by research and practice on fair algorithmic hiring, have the potential to influence the industry positively and profoundly.  

\namedpar{Limitations}. Most of the fairness approaches developed so far are restricted to proxy reduction or removal, neglecting a wealth of solutions developed by the algorithmic fairness community. This is most likely due to concerns of infringing regulation on disparate treatment and direct discrimination. Furthermore, special categories of personal data are \wasnew{often} lacking and difficult to process\wasnew{, particularly in the EU} under the current data protection law. Therefore, it is \wasreplace{impossible to assess hiring practices for vulnerable populations}{difficult to assess hiring practices for certain vulnerable populations}. Data protection law restricts these analyses and should consider exceptions for algorithmic fairness, as now suggested in the EU AI Act.

\namedpar{Recommendations}. Algorithmic fairness can conflict with privacy and non-discrimination doctrine. The exact contours of legally compliant algorithmic fairness remain contested. The EU AI Act may offer a (limited) solution by allowing certain types of sensitive data processing to remove biases in high-risk scenarios. This guidance should be expanded to other areas and jurisdictions. We advocate for further multidisciplinary research on this topic, studying technical solutions and legal frameworks to reconcile these principles in light of their trade-offs. Promising technological approaches include multiparty computation \citep{kilbertus2018blind}, sample-level estimators \citep{fabris2023measuring}, and noise injection mechanisms \citep{juarez2023you}.

\section{Conclusions}
\label{sec:concl}

The social, technological, and legal landscape around algorithmic hiring is rapidly evolving; algorithmic fairness has  become a necessary component for both business-as-usual product development and frontier research. 
Practitioners and researchers in this field must understand bias conducive factors, leveraging contextualized measures carried out on appropriate data to deploy suitable bias mitigation strategies. Multidisciplinary work at the intersection with legal scholarship is especially critical to implement and guide policy by defining technically achievable desiderata. Only a contextualized and balanced understanding of fair algorithmic hiring can guide research and practice to avoid the pitfalls of legitimizing questionable applications with misguided analyses and to reap truly shared benefits for society.


\begin{acks}
We are indebted to many researchers and practitioners for advice on this work, including Anisha Nadkarni, Anna Via, Carlos Castillo, Clara Rus, Didac Fortuny Almiñana, Feng Lu, Ilir Kola, Justine Devos, Marc Serra Vidal, and Volodymyr Medentsiy. 

\noindent This work is supported by the \href{https://findhr.eu/}{FINDHR} project, Horizon Europe grant agreement ID: 101070212 and by the Alexander von Humboldt Foundation.
\end{acks}

\newpage

\DeclareRobustCommand\EEOClongname{ - US Equal Employment Opportunity Commission}
\DeclareRobustCommand\UNDPlongname{ - United Nations Development Programme}
\bibliographystyle{ACM-Reference-Format}
\bibliography{biblio}

\appendix

\section{Systematic Review Methodology}
\label{sec:lit_rev}
This article is an interdisciplinary survey aimed at informing researchers and practitioners interested in fairness and bias in algorithmic hiring. We focused on a Computer Science (CS) perspective while summarizing key topics from Human Resource Management, Industrial and Organizational Psychology, Philosophy and Law with mixed methods, including the analysis of influential technical reports, industry white papers and legal literature. Three sections of this work, summarizing measures, mitigation strategies, and data (Sections \ref{sec:measures}-\ref{sec:data}), result from a systematic literature review summarized below.

\begin{enumerate}
    \item To ensure a broad coverage of the scientific literature centered on CS, we leverage three scholarly search engines: IEEE Xplore, ACM Digital Library, and Google Scholar.
    \item We use the query \texttt{algorithmic} $\wedge$ \texttt{hiring} $\wedge$ \texttt{fairness} on article titles, where each term is expanded as follows:
    \begin{itemize}
        \item \texttt{algorithmic}:  algorithm* $\vee$ AI $\vee$ search* $\vee$ recommend* $\vee$ rank* $\vee$ screen* $\vee$ retriev* 
        \item \texttt{hiring}: hir* $\vee$ recruit* $\vee$ candidate* $\vee$ job* $\vee$ work* $\vee$ resum* $\vee$ CV $\vee$ interview* $\vee$ eval* $\vee$ appraisal. The last two terms target the evaluation stage. 
        \item \texttt{fairness}: *bias* $\vee$ *ethic* $\vee$ *fair* $\vee$ discriminat* $\vee$ *equit* $\vee$ *equal* $\vee$ *parit* $\vee$ *symmetr* $\vee$ gap
    \end{itemize}
    \item To ensure high recall, we consider the top 100 results. To guarantee precision, we manually analyze each article and only select the ones that treat fairness in algorithmic hiring from a quantitative perspective, i.e. performing a fairness audit or introducing a novel method. To exemplify, we discard articles on related yet different topics, such as freelancing \citep{hannak2017bias}, focused on human perceptions \citep{lavanchy2023applicants}, on qualitative aspects \citep{vandenbroek2019hiring}, or mitigating biases to improve accuracy without any fairness consideration \citep{chen2019correcting}.
    \item We take additional steps to further improve recall. For each included article, we perform forward and backward snowballing \citep{wohlin2014guidelines}, pre-filtering article titles with a ``\texttt{hiring} $\wedge$ \texttt{fairness}'' query and applying the inclusion criteria in (3). Finally, we consider articles and datasets presented in related surveys \citep{fabris2022algorithmic,mashayekhi2022challenge,zehlike2023fairness,fabris2022tackling} or published at dedicated venues,\footnote{Workshop on Recommender Systems for Human Resources \url{https://recsyshr.aau.dk/}} finding one additional dataset (IBM HR Analytics) and no additional article.
\end{enumerate}

\end{document}